\documentclass[pdftex,dvipdfm,epsf]{iopart}
\eqnobysec
\usepackage{iopams,amsfonts,rotating,graphicx,fancybox,fancyhdr,dsfont}
\usepackage{eucal}
\usepackage{verbatim}
\usepackage{appendix}

\newcommand{\mkbox}[3]{\hbox{\vrule
      \vbox to  #1{\hrule \vss
                  \hbox to #2{\hss#3\hss}\vss
                  \hrule}\vrule}}
\newcommand{\be}{\begin{equation}}
\newcommand{\ee}{\end{equation}}
\newcommand{\bea}{\begin{eqnarray}}
\newcommand{\eea}{\end{eqnarray}}

\newcommand{\la}{\lambda}

\def\Tr{\mathrm{Tr}}

\newcommand{\newc}{\newcommand}
\newc{\beq}{\begin{equation}}
\newc{\eeq}{\end{equation}}
\newc{\kt}{\rangle}
\newc{\bra}{\langle}
\newc{\beqa}{\begin{eqnarray}}
\newc{\eeqa}{\end{eqnarray}}
\newc{\pr}{\prime}
\newc{\longra}{\longrightarrow}
\newc{\ot}{\otimes}
\newc{\rarrow}{\rightarrow}
\newc{\h}{\hat}
\newc{\bom}{\boldmath}
\newc{\btd}{\bigtriangledown}
\newc{\al}{\alpha}
\newc{\ld}{\lambda}
\newc{\ldmin}{\lambda_{\rm min}}
\newc{\sg}{\sigma}
\newc{\p}{\psi}
\newc{\eps}{\epsilon}
\newc{\om}{\omega}
\newc{\mb}{\mbox}
\newc{\tm}{\times}
\newc{\hu}{\hat{u}}
\newc{\hv}{\hat{v}}
\newc{\hk}{\hat{K}}
\newc{\ra}{\rightarrow}
\newc{\non}{\nonumber}
\newc{\ul}{\underline}
\newc{\hs}{\hspace}
\newc{\longla}{\longleftarrow}
\newc{\ts}{\textstyle}
\newc{\f}{\frac}
\newc{\df}{\dfrac}
\newc{\ovl}{\overline}
\newc{\bc}{\begin{center}}
\newc{\ec}{\end{center}}
\newc{\dg}{\dagger}
\newc{\prh}{\mbox{PR}_H}
\newc{\prq}{\mbox{PR}_q}

\begin{document}

\title{Compact smallest eigenvalue expressions in Wishart- 
Laguerre ensembles with or without fixed-trace}

\author{Gernot Akemann}
\address{Department of Physics, Bielefeld University, Postfach 100131, D-33501 Bielefeld, Germany}
\ead{akemann@physik.uni-bielefeld.de}

\author{Pierpaolo Vivo}
\address{Abdus Salam International Centre for
Theoretical Physics, Strada Costiera 11, 34151 Trieste, Italy}
\ead{pvivo@ictp.it}
\begin{abstract}
The degree of entanglement of random pure states in bipartite quantum systems can be estimated from the distribution of the extreme Schmidt eigenvalues. 
For a bipartition of size $M\geq N$, these are distributed according to a Wishart-Laguerre ensemble (\textsf{WL}) of random matrices
of size $N\times M$, with a fixed-trace constraint. We first compute the distribution and moments of the smallest eigenvalue in the fixed trace orthogonal \textsf{WL} ensemble
for arbitrary $M\geq N$. Our method is based on a Laplace inversion of the recursive results for the corresponding orthogonal \textsf{WL} ensemble by Edelman. 
Explicit examples are given for fixed $N$ and $M$, generalizing and simplifying earlier results.
In the microscopic large-$N$ limit with $M-N$ fixed, the orthogonal and unitary \textsf{WL} distributions exhibit universality after a suitable rescaling and 
are therefore independent of the constraint. We prove that very recent results given in terms of hypergeometric functions of matrix argument are equivalent to 
more explicit expressions in terms of a Pfaffian or determinant of Bessel functions.
While the latter were mostly known from the random matrix literature on the \textsf{QCD} Dirac operator spectrum, we also derive some new results in the orthogonal symmetry class. 

\end{abstract}

\newpage
\tableofcontents
\newpage

\maketitle

\section{Introduction}

The theory of random matrices finds applications to the most diverse physical situations, and Wigner and Dyson are usually referred to as the pioneers of this field for
their works on nuclear spectra. However, many years before their seminal papers, John Wishart had already introduced random covariance matrices in his studies on multivariate
populations \cite{Wishart}. The Wishart ensemble (also named Laguerre or chiral, and called \textsf{WL} hereafter) contains random $N\times N$ covariance matrices of the form 
$\mathcal{W}=\mathcal{X}^\dagger\mathcal{X}$ where $\mathcal{X}$ is a $M\times N$ matrix with i.i.d. Gaussian entries (real, complex or quaternions labelled by the Dyson index $\beta=1,2,4$ respectively). 
The joint probability density function
(\textsf{jpdf}) of its $N$ non-negative eigenvalues $\{\lambda_i\}$ is given in Eq. (\ref{wishart11}). A related ensemble of random matrices is the fixed-trace Wishart-Laguerre (\textsf{FTWL}),
which contains \textsf{WL} matrices with a prescribed trace $\Tr(\mathcal{W})=t>0$ and whose eigenvalue \textsf{jpdf} is given in (\ref{jpdsch}). Recent results
about the density of eigenvalues of \textsf{FTWL} matrices can be found in \cite{densitybeta2,adachi,vivodensity}. The study of fixed and bounded trace ensembles of non-chiral random matrices
has however a longer history \cite{Mehta}. In \cite{ACMV,Delsemi} the spectral density of the related, fixed trace Gaussian Unitary Ensemble was computed explicitly for finite $N$, while for works on non-chiral fixed trace $\beta$-ensembles we refer to \cite{LeCFTbeta,LZFTbeta}.

Two recent applications of \textsf{WL} and \textsf{FTWL} ensembles that motivate our study are effective theories of Quantum Chromodynamics (\textsf{QCD}) and \textsf{QCD}-like theories, where we refer to \cite{Jac}
for a most recent review and references, and the statistical theory of entangled random pure states in bipartite systems, see \cite{majreview} for an excellent review, respectively.

Entanglement is indeed one of the most distinctive features of quantum systems, and it is a crucial resource in quantum computation issues, as performances of quantum computers
will heavily rely on the possibility of producing states with large entanglement \cite{NeilsenBook,PeresBook}. In order to quantify the degree of entanglement of a given quantum state,
it is useful to introduce an unbiased benchmark of states with the lowest degree of built-in information: pure bipartite systems (defined below) with a Hamiltonian of size $M\cdot N$
constitute a typical example
where well-behaved entanglement quantifiers can be defined, such as the von-Neumann or R\'enyi entropies of either subsystem \cite{PeresBook}, the so
called concurrence for two-qubit systems~\cite{Wootters} or other entanglement monotones \cite{cappellini,gour}.

The main focus of this paper is on the cumulative distribution and density of the smallest \textsf{FTWL} eigenvalue and its relation to the corresponding quantity for the 
unconstrained \textsf{WL} ensemble: these two distributions arise naturally in the two settings described above
(entanglement and \textsf{QCD}, respectively). The upshot is that the smallest Schmidt eigenvalue (a relevant quantity in the entanglement setting discussed below) is precisely 
given by the smallest eigenvalue of 
one of the \textsf{FTWL} ensembles with $\Tr(\mathcal{W})=t=1$.
Which of these ensembles to choose depends on the global symmetry of the quantum system, and we will report on the orthogonal ($\beta=1$) and unitary ($\beta=2$)  symmetry classes, 
but not on the symplectic ($\beta=4$) class.
Another application of the smallest \textsf{FTWL} eigenvalue is
related to the Demmel condition number \cite{demmel}, which quantifies how hard certain numerical problems about matrix inversion are.

The results of this paper can be grouped into two parts:
\begin{enumerate}
\item  First, we study the full distribution of the smallest eigenvalue of \textsf{FTWL}
orthogonal ensemble ($\beta=1$) at finite $N,M$. 
\item Next, we take the large-$N,M$ limit of the smallest eigenvalue with $\nu=M-N\geq 0$ fixed (the so-called \emph{microscopic} limit) for both $\beta=1,2$. 
\end{enumerate}
Let us summarize first which results were previously known for both \textsf{FTWL} and unconstrained \textsf{WL}:
\begin{itemize}
\item For \textsf{FTWL} ensemble at finite $N,M$, the full distribution of the smallest eigenvalue was derived for $\beta=1,2$
and $M=N$ in \cite{majbohi}\footnote{In studying the distribution of the Demmel
condition number, Edelman \cite{edeldemmel} has derived the same formulae in a
completely different setting. An exact mapping between the two
problems is possible, but nontrivial \cite{ZS,ZyckBook}.}, and for $\beta=4$ in \cite{majreview}. In \cite{majbohi}, a conjecture by Znidaric \cite{Znd} was proven for the first moment.
For general real $\beta>0$ and $m\in\mathbb{N}$, where $m=(\beta/2)(M-N+1)-1$, Chen \emph{et al.} \cite{chen} reported formal expressions for the smallest eigenvalue cumulative distribution
and density in terms of cumbersome sums of Jack polynomials. More explicit expressions are given for the case $\beta=2$ in terms of finite sums.
\item For \textsf{WL} ensemble at finite $N,M$, the first results for the smallest eigenvalues of the orthogonal \textsf{WL} ensemble go back to Edelman \cite{Edel,Edelrec} 
who gave a recursive prescription.
Curiously, analogous recurrence relations for $\beta=2,4$ have not been worked out to date, while for $\beta=2$ a closed expression exists in terms of determinants \cite{forrhughes}.
These results
were extended by Forrester \cite{Forrester1st} to the case of real $\beta>0$ for special values of $\nu=M-N$. Forrester again waived some limitations on $\nu$ and further generalized 
these results \cite{forrwaive}. The most up-to-date and general formula for the smallest eigenvalue distribution involves hypergeometric function of matrix argument
(\textsf{HFMA}), see eq. 2.13b in \cite{forrwaive}, as well as our Appendix \ref{hypA}. Note that for $\beta=1$ this requires
$\nu$ to be odd. For $\beta=1,4$ expressions in terms of Pfaffians were reported \cite{nagaoforr}, with the same restriction for $\beta=1$, and
 for $\beta=4$ the values of $\nu>0$ are half-integers.
\end{itemize}
The next natural step is to take the large $N,M$ limit for the smallest eigenvalue. For real $\beta>0$ keeping $\nu$ fixed, this was done in \cite{forrwaive} for the \textsf{WL} 
and subsequently in \cite{chen}
for \textsf{FTWL} ensembles, expressing them in terms of \textsf{HFMA}. It turned out that in this large $N$ limit (after a suitable rescaling) the fixed trace constraint does not matter, and in that sense the results are 
universal (for a different limit, namely $c=N/M$ fixed, see the recent paper \cite{castillo}).

The issue of universality in the fixed (or bounded) trace ensembles had
been addressed earlier in the literature, mainly for the non-chiral ensembles.
It was shown in \cite{ACMV} that in the \emph{macroscopic} large $N$ limit where the oscillations of the density are smoothed, the semi-circle or its generalizations 
can be matched in the constrained and unconstrained 
ensembles. In contrast, the higher order connected
correlators become non-universal when considering non-Gaussian generalized \textsf{WL} ensembles, with a
polynomial potential in ${\mathcal W}$ in the exponent. 
On the other hand, in the \emph{microscopic} limit in the bulk of the spectrum, where the oscillatory behavior of the density is zoomed into, the constraint was found to be immaterial and 
the universality of the sine-kernel was established in
\cite{AV}. The same result was provided later in a mathematically more rigorous way for non-invariant generalizations of \textsf{WL} in \cite{Goetze,Liu}.

Because of this universality (i.e. constraint-independence), it is extremely useful to recall the large $N$ results derived
independently for the smallest eigenvalue distributions in the application of unconstrained \textsf{WL} ensembles to \textsf{QCD} \cite{WGW,DNW,DN98}.
Here also non-Gaussian generalisations of \textsf{WL} were considered, and determinantal or Pfaffian expressions of Bessel functions were derived 
for $\beta=2$ \cite{DNW} and $\beta=1$ \cite{DN98}, respectively.
In fact much more general results were derived there, including an arbitrary number of characteristic polynomials of random matrices (so-called mass terms) in the weight function.
In the second part of this paper we will show how these results in the \textsf{QCD} literature and the aforementioned ones in terms of \textsf{HFMA} are related.

The presentation of the paper is organized as follows. \\
In order to be self-contained in section \ref{setting} we provide some background material about the relation between bipartite entanglement and \textsf{FTWL} 
(subsection \ref{entangle}). In subsection \ref{WLdefs} we give definitions and notations used for \textsf{WL} and \textsf{FTWL} ensembles. The reader familiar to either or both of these topics 
may skip the corresponding subsection(s).

In the next section \ref{NewWLFT} we derive our new results for \textsf{FTWL} ensembles at $\beta=1$ for arbitrary finite $N$ and $M$, both for odd and even $\nu$. There we briefly recall the results for standard \textsf{WL} by Edelman on which we heavily rely. This section also contains our results for arbitrary moments in subsection \ref{MomentsN} and numerical checks in \ref{numcheck}.

Section \ref{equiv} brings us to the universal microscopic limit for
fixed $\nu=M-N$. It provides equivalence proofs between the known results from the \textsf{QCD} Dirac spectrum literature,
given in subsection \ref{besselsec},
and the very recent results in terms of \textsf{HFMA}, given in subsection \ref{sectHFMA}. 
In subsection \ref{scaledsubnu} we explicitly compute the large $N$ limit of the scaled smallest eigenvalue for $\beta=1$ and $\nu=0,2$ from their finite $N$
expressions and confirm the universality results. For
$\beta=2$ the above mentioned equivalence is established in subsection \ref{beta2}, and for $\beta=1$ in
subsection \ref{beta1}, including new results for the cumulative distribution and for
even $\nu=2$.
We also report on the general scaling for moments in subsection \ref{Momentsoo} before concluding.

Some technical details about the definition of \textsf{HFMA}
and a universality check for $\beta=1$ and $\nu=0$ are deferred to the appendices.

\section{Bipartite entanglement and \textsf{FTWL} ensembles}\label{setting}

\subsection{Bipartite entanglement}\label{entangle}

Let ${\cal
H}^{(NM)}$ be a $N\cdot M$-dimensional Hilbert space which is bipartite as ${\cal H}^{(NM)}={\cal H}^{(N)}_A \otimes {\cal H}^{(M)}_B$, where $N\le M$ without loss of generality.
For example, $A$ may be taken as a given
subsystem (say a set of spins) and $B$ may represent the environment (e.g., a
heat bath).
Let $\{|i^A\kt\}$ and $\{|\alpha^B\kt\}$ be two complete bases of ${\cal H}^{(N)}_A $ and ${\cal H}^{(M)}_B$ respectively. Then, any quantum state $|\psi\kt$ 
of the composite system can be decomposed as:
\begin{equation}\label{psi111}
|\psi\kt= \sum_{i=1}^N\sum_{\alpha=1}^M x_{i,\alpha}\,
|i^A\kt\otimes |\alpha^B\kt
\end{equation}
where the coefficients $x_{i,\alpha}$'s form the entries of a rectangular $(N\times M)$ matrix $\mathcal{X}$.

In the following, we will consider entangled random pure states $|\psi\kt$. This means that:
\begin{enumerate}
\item $|\psi\kt$ \emph{cannot} be expressed as a direct
product
of two states belonging to the two subsystems $A$ and $B$. 
\item the expansion coefficients
$x_{i,\alpha}$ are random variables drawn from a certain probability distribution (see below).
\item the density matrix of the composite system is simply given by $\rho=|\psi\kt \bra \psi|$ with the constraint ${\rm Tr}[\rho]=1$, or equivalently $\bra \psi|\psi\kt=1$.
\end{enumerate}

We will not consider statistically mixed states here, and we refer to \cite{osipov} and references therein for recent results on mixed states.

The density matrix $\rho$ of a quantum state is a very important quantity as it allows to compute expectation values of observables. For an entangled pure state $|\psi\kt$ of a bipartite quantum system 
it can then be straightforwardly written as:
\beq
\rho = \sum_{i,\alpha}\sum_{j,\beta} x_{i,\alpha}\, x_{j,\beta}^*\, |i^A\kt\bra j^A|\otimes
|\alpha^B\kt
\bra\beta^B|,
\label{dem2}
\eeq
where the Roman indices $i$ and $j$ run from $1$ to $N$ and the Greek indices $\alpha$ and
$\beta$ run from $1$ to $M$.

In some applications, it is useful to separate the contribution of the subsystem $A$ under consideration from the environment $B$. Expectation values of observables $\mathcal{O}_A$ on the system $A$ can be obtained by ``tracing out'' the environmental degrees of freedom (i.e., those of subsystem $B$) and defining the \emph{reduced} density matrix $\rho_A= {\rm Tr}_B[\rho]$ as:
\beq
\rho_A = {\rm Tr}_B[\rho]=\sum_{\alpha=1}^M \bra \alpha^B|\rho|\alpha^B\kt.
\label{rdm1}
\eeq
Using the expansion in Eq. (\ref{dem2}) one gets
\beq
\rho_A = \sum_{i,j=1}^N \sum_{\alpha=1}^M x_{i,\alpha}\, x_{j,\alpha}^*\, |i^A\kt\bra
j^A|=\sum_{i,j=1}^N W_{ij} |i^A\kt\bra j^A|
\label{rdm2}
\eeq
where $W_{ij}$'s are the entries of the $N\times N$ covariance matrix $\mathcal{W}=\mathcal{X} \mathcal{X}^{\dagger}$.

Proceeding further, one could also obtain the reduced density matrix $\rho_B={\rm Tr}_A[\rho]$ of the
subsystem $B$ in terms of the $M\times M$ matrix $\mathcal{W}^\prime=\mathcal{X}^\dagger \mathcal{X}$ and find that $\mathcal{W}$ and $\mathcal{W}^\prime$ share
the same set of nonzero (positive) real eigenvalues $\{\lambda_1,\lambda_2,\ldots,\lambda_N\}$, called {\em Schmidt} eigenvalues.

In the basis of eigenvectors of $\mathcal{W}$, one can express $\rho_A$ as
\beq
\rho_A= \sum_{i=1}^N \lambda_i \, |\ld^A_i\kt\, \bra \ld^A_i|
\label{diagA}
\eeq
where $|\ld^A_i \kt$'s are the normalized eigenvectors of $\mathcal{W}=\mathcal{X}\mathcal{X}^{\dagger}$.
The original composite state $|\psi\kt$ in this diagonal basis reads:
\beq
|\psi\kt = \sum_{i=1}^{N} \sqrt{\ld_i}\, |\ld_i^A\kt \otimes |\ld^B_i \kt.
\label{Sch1}
\eeq
Eq. (\ref{Sch1}) is known as the Schmidt decomposition, and the
normalization
condition $\bra \psi|\psi\kt=1$, or equivalently ${\rm Tr}[\rho_A]=1$, imposes
a constraint on the eigenvalues, $\sum_{i=1}^N \ld_i=1$.

In the Schmidt decomposition (\ref{Sch1}), each state $|\ld_i^A\kt \otimes|\ld^B_i \kt$ is separable, but their linear combination
 $|\psi\kt$ (depending on the set of Schmidt eigenvalues) cannot, in general,
be written as a direct product $|\psi\kt= |\phi^A\kt \otimes |\phi^B\kt$ of two states of the
respective subsystems, i.e. it is entangled. Knowledge of the Schmidt eigenvalues $\{\lambda_1,\lambda_2,\ldots, \lambda_N\}$ of the matrix $\mathcal{W}$
is therefore essential in providing information
about how entangled a pure state is.

For \emph{random} pure states, the expansion coefficients in eq. (\ref{psi111})
can be typically drawn from an unbiased (so called \emph{Hilbert-Schmidt}) distribution
\begin{equation}\label{HS}
{\rm Prob}[\mathcal{X}]\propto \delta\left( {\rm Tr}(\mathcal{X} \mathcal{X}^{\dagger})-1\right).
\end{equation}
The meaning of eq.
(\ref{HS}) is clear: all normalized density matrices are sampled with equal probability, which corresponds to having minimal \emph{a priori} information about the quantum state under consideration. This in turn induces nontrivial correlations among the Schmidt eigenvalues (which are now real random variables between $0$ and $1$ whose sum is $1$) and makes the investigation of several statistical quantities about such states quite interesting. The \textsf{jpdf} of Schmidt eigenvalues for a Hilbert-Schmidt distribution of coefficients was derived in \cite{LP} and turns out to be exactly of the \textsf{FTWL} form (\ref{jpdsch})
with $t=1$, where
the Dyson index $\beta=1,2$ corresponds respectively to real and complex
$\mathcal{X}$ matrices\footnote{These two cases in turn correspond to quantum systems whose Hamiltonians preserve ($\beta=1$) or break ($\beta=2$) time-reversal symmetry.}. The delta constraint there indeed guarantees a proper normalization of the reduced density matrix ${\rm Tr}[\rho_A]=1$.

Why is the smallest Schmidt eigenvalue distribution interesting at all?
We first note that due to the constraint $\sum_{i=1}^N \ld_i=1$ and the fact that
all eigenvalues are nonnegative, it follows that $1/N\le \ld_{\rm max}\le 1$
and $0\le \ld_{\rm min} \le 1/N$. Now consider the following
limiting situations. Suppose that the smallest eigenvalue $\lambda_{\rm min}=\min_i\{\lambda_i\}$ takes its maximum allowed value $1/N$. Then
it follows immediately that all the remaining $(N-1)$ eigenvalues must be also identically equal to $\lambda_i=1/N$. In this situation eq. (\ref{Sch1}) tells us that $|\psi\kt$ is
maximally entangled. On the other hand, if $\lambda_{\rm min}=0$ (i.e., it takes its lowest allowed value) or close to $0$,
while this will not provide much information about the degree of entanglement of $|\psi\kt$, it actually tells us that one component in the Schmidt decomposition (\ref{Sch1})
can be safely ignored. In other words, the 'effective' dimension of the Hilbert space $\mathcal{H}_A$ has been reduced from $N$ to $N-1$. The proximity
of the smallest eigenvalue to zero, therefore, gives information about the efficiency of this dimensional reduction process.

For more references on entangled random pure states we refer to \cite{majreview}.

\subsection{\textsf{WL} Ensembles with and without fixed trace}\label{WLdefs}

The \textsf{jpdf} of non-negative eigenvalues of the unconstrained \textsf{WL} ensemble is given by:
\beq
\fl\mathcal{P}^{(\mathrm{WL})}(\lambda_1,\ldots,\lambda_N) =K_{N,M}^{(\beta)}\, \exp\left[-\frac{\beta}{2}\sum_{i=1}^N\lambda_i\right]
\, \prod_{i=1}^N \lambda_i^{\frac{\beta}{2}(1+M-N)-1}\, |\Delta(\vec{\lambda})|^\beta
\label{wishart11}
\eeq
where $K_{N,M}^{(\beta)}$ is a known normalization constant: 
\beq
\label{KWL}
K_{N,M}^{(\beta)}=\left(\frac{\beta}{2}\right)^{\frac{\beta}{2}NM}\prod_{j=1}^N\frac{\Gamma\left(1+\frac{\beta}{2}\right)}{\Gamma\left(1+\frac{\beta}{2}j\right)\Gamma\left(\frac{\beta}{2}(M-N+j)\right)}
\ ,
\eeq
and $\Delta(\vec{\lambda})=\prod_{j<k}(\lambda_j-\lambda_k)$ is the Vandermonde determinant.

On the other hand, the \textsf{jpdf} of the eigenvalues $\lambda_i\in [0,t]$ of the \textsf{FTWL} ensemble is given by:
\begin{equation}\label{jpdsch}
\fl\mathcal{P}^{(\mathrm{FT})}(\lambda_1,\ldots,\lambda_N;t)
=C_{N,M}^{(\beta)}(t)\delta\left(\sum_{i=1}^N\lambda_i -t\right)\prod_{i=1}^N\lambda_i^{\frac{\beta}{2}(M-N+1)-1}|\Delta(\vec{\lambda})|^\beta.
\end{equation}
For $t=1$ it coincides with the distribution of Schmidt eigenvalues, where
the delta function guarantees that ${\rm Tr}[\rho]=\sum_{i=1}^N\la_i=t$. The normalization constant for $t=1$ reads \cite{ZS}:
\beq
\label{Cft}
C_{N,M}^{(\beta)}\equiv C_{N,M}^{(\beta)}(1)=\frac{\Gamma(MN\beta/2)(\Gamma(1+\beta/2))^N}{\prod_{j=0}^{N-1}\Gamma((M-j)\beta/2)\Gamma(1+(N-j)\beta/2)}\ .
\eeq

The presence of a fixed-trace constraint is known to have crucial consequences on (connected) spectral correlation functions, both for finite-$N$ and in the macroscopic large-$N$ limit \cite{ACMV,ASOS}.
However, in the microscopic large-$N$ limit the correlations become universal, and we shall exploit this in the next section.
Let us first introduce the crucial quantities for this paper, the cumulative distributions
$q_{N,\nu}(x)=\mathrm{Prob}[\lambda_{\mathrm{min}}>x]$ (also called gap-probabilities),
and the corresponding densities for the smallest eigenvalue of both ensembles:
\begin{eqnarray}
\fl q_{N,\nu}^{(\mathrm{FT})}(x;t) &=C_{N,M}^{(\beta)}(t)\int_{[x,\infty]^N}\prod_{i=1}^N d\lambda_i\delta\left(\sum_{i=1}^N\lambda_i -t\right)\prod_{i=1}^N\lambda_i^{\frac{\beta}{2}(\nu+1)-1} |\Delta(\vec{\lambda})|^\beta\ ,\\
\fl p_{N,\nu}^{(\mathrm{FT})}(x;t) &=-\frac{\partial}{\partial x}q_{N,\nu}^{(\mathrm{FT})}(x;t)\ ,
\label{Qpdel}\\
\fl q_{N,\nu}^{(\mathrm{WL})}(x) &=K_{N,M}^{(\beta)}\int_{[x,\infty]^N}\prod_{i=1}^N d\lambda_i e^{-\frac{\beta}{2}\sum_{i=1}^N\lambda_i}\prod_{i=1}^N\lambda_i^{\frac{\beta}{2}(\nu+1)-1}|\Delta(\vec{\lambda})|^\beta\ ,\\
\fl p_{N,\nu}^{(\mathrm{WL})}(x) &=-\frac{d}{dx}q_{N,\nu}^{(\mathrm{WL})}(x)\ ,
\label{QpWL}
\end{eqnarray}
where the superscript distinguishes fixed-trace $(\mathrm{FT})$  and ordinary \textsf{WL}.
We have the following normalization conditions:
\begin{eqnarray}
\int_0^\infty dx\ p_{N,\nu}^{(\mathrm{WL})}(x) =\int_0^{t/N} dx\ p_{N,\nu}^{(\mathrm{FT})}(x;t) =1\ ,\\
q_{N,\nu}^{(\mathrm{WL})}(0) = q_{N,\nu}^{(\mathrm{FT})}(0;t) = 1 \ .
\label{normN}
\end{eqnarray}
The $\ell$th moments of the smallest Schmidt and \textsf{WL} eigenvalue 
are then given by
\begin{eqnarray}
\label{momentdef}
\langle \lambda_{\mathrm{min}}^\ell\rangle_{N,\nu,\beta}^{(\mathrm{FT})}&\equiv&\int_0^{1/N} dx\ x^\ell p_{N,\nu}^{(\mathrm{FT})}(x)\ ,
\\
\label{momentdef2}\langle \lambda_{\mathrm{min}}^\ell\rangle_{N,\nu,\beta}^{(\mathrm{WL})}&\equiv&\int_0^\infty dx\ x^\ell p_{N,\nu}^{(\mathrm{WL})}(x)\ .
\end{eqnarray}
where we define $p_{N,\nu}^{(\mathrm{FT})}(x):=p_{N,\nu}^{(\mathrm{FT})}(x;1)$ for later convenience.
All density correlation functions and thus also the smallest eigenvalue distribution in ensembles with and without fixed trace
can be related via an inverse  Laplace transform, see e.g. \cite{AV}
\footnote{For this reason we have kept the fixed trace to be a free parameter $t$ in the \textsf{jpdf} (\ref{jpdsch}).}. For the smallest eigenvalue distribution this relation reads:
\begin{eqnarray}
\fl\mathcal{L}[{p}_{N,\nu}^{(\mathrm{FT})}(x;t)](s)&=&\int_0^\infty dt\ {p}_{N,\nu}^{(\mathrm{FT})}(x;t) e^{-st}\non\\
\fl&=&\frac{C_{N,M}^{(\beta)}}{K_{N,M}^{(\beta)}}\left(\frac{1}{2s}\right)^{-1+MN\frac{\beta}{2}
+\frac{(1-\beta)}{2}N(N-1)}
p_{N,\nu}^{(\mathrm{WL})}(2sx)\ .
\label{mainlapp}
\end{eqnarray}

This is the main technical tool we use in the next section.
For finite $N,M$ we are going to use known explicit expressions for \textsf{WL}
smallest eigenvalue densities
in order to derive the sought density for \textsf{FTWL} via the inverse Laplace transform of eq. (\ref{mainlapp}).

\section{New results for \textsf{FTWL} ensemble at finite $N,M$ and $\beta=1$}
\label{NewWLFT}

In this section we will focus on the $\beta=1$ ensemble only. Our strategy is
very simple. We will take the known results for the standard $\beta=1$ \textsf{WL}
ensemble at any finite $N,M$ derived by Edelman \cite{Edelrec}  
and invert the Laplace transform in eq. (\ref{mainlapp}).
The same strategy was followed by Chen \emph{et al.} \cite{chen}, however, they applied the inverse Laplace transform directly to the \textsf{HFMA} result by Forrester \cite{forrwaive} 
valid for real $\beta>0$.
This led to expressions that are formally exact, but not very transparent or manageable. Furthermore,
in contrast to \cite{chen,forrwaive} we will not be restricted to odd $\nu=M-N$  for $\beta=1$.
The special case for $\nu=0$ was derived independently in \cite{edeldemmel} and \cite{majbohi}.

\subsection{Odd $\nu=M-N$}
\label{sectnuodd}

We begin with the simpler case.
Edelman \cite{Edelrec} states that the smallest eigenvalue distribution for a \textsf{WL} ensemble with $\beta=1$ and $\nu=M-N$ odd
can be written as follows:\footnote{Compared to the notation in \cite{Edelrec} we spell out the parameter $\rho$ explicitly contained in $c_{N,\nu}$ there.}
\begin{equation}\label{pnuedel}
p_{N,\nu}^{(\mathrm{WL})}(x) = 2^{\frac{N}{2}-1} c_{N,\nu} x^{(\nu-1)/2}e^{-Nx/2}h_{N,\nu}(x)
\end{equation}
where $h_{N,\nu}(x)$ is a polynomial of degree $(N-1)(\nu-1)/2$ and $c_{N,\nu} $ is the following constant:
\begin{equation}
 c_{N,\nu}=\frac{N 2^{-N\nu/2}\Gamma((N+1)/2)}{\sqrt{\pi}}\prod_{j=1}^{\nu}\frac{\Gamma(j/2)}{\Gamma((N+j)/2)}\ .
\label{cNnu}
\end{equation}

As such, $h_{N,\nu}(x)$ can be written as:
\begin{equation}\label{hedel}
h_{N,\nu}(x)=\sum_{k=0}^{\frac{(N-1)(\nu-1)}{2}} \mathfrak{q}_k x^k
\end{equation}
with some rational coefficients $\mathfrak{q}_k$, which depend also on
$N,\nu$. The polynomial $h_{N,\nu}$ is 
determined via a simple recursion relation explicitly given in \cite{Edelrec}.
In addition, a simple Mathematica$\textsuperscript{\textregistered}$ code is given in Appendix A of \cite{Edelrec} to generate the
$h_{N,\nu}(x)$. The first two polynomials read:
\be
h_{N,\nu=1}(x)=1\ ,\ \ h_{N,\nu=3}(x)= \frac{2^N\Gamma\left(\frac{N}{2}+1\right)\Gamma\left(\frac{N+3}{2}\right)}{N(N+1)\Gamma\left(\frac32\right)}
L_{N-1}^{(3)}(-x) \ ,
\label{hnu13}
\ee
and so on using the given recursion. In (\ref{hnu13}), $L_n^{(\alpha)}(x)$ is a generalized Laguerre polynomial\footnote{Note a typo in Lemma 4.5. in 
\cite{Edelrec} in the representation of $L_n^{(\alpha)}(-x)$.}.

Now, it is easy to observe
that (no matter what the coefficients $\mathfrak{q}_k$ are) the form (\ref{pnuedel}) lends itself to a very friendly Laplace inversion.
More precisely, take the inverse Laplace transform of the fundamental relation (\ref{mainlapp}) after inserting eq. (\ref{hedel}) into (\ref{pnuedel}). One obtains:
\begin{eqnarray}
\nonumber{p}_{N,\nu}^{(\mathrm{FT})}(x;t) &=&\frac{C_{N,M}^{(1)}}{K_{N,M}^{(1)}} 2^{\frac{N}{2}-1}c_{N,\nu}\sum_{k=0}^{\frac{(N-1)(\nu-1)}{2}} \mathfrak{q}_k x^{k+(\nu-1)/2}\times\\
&&\times\mathcal{L}^{-1}\left[
(2s)^{(2k+\nu+1-N(N+\nu))/2}e^{-Nsx}\ .
\right](t)
\end{eqnarray}
Computing the inverse Laplace transform and setting $t=1$, one gets the final general formula:
\begin{eqnarray}
\fl\nonumber  p_{N,\nu}^{(\mathrm{FT})}(x) &=&\frac{C_{N,M}^{(1)}}{K_{N,M}^{(1)}} 
2^{\frac{N}{2}-1}c_{N,\nu}\sum_{k=0}^{\frac{(N-1)(\nu-1)}{2}} \frac{\mathfrak{q}_k\ 2^{(2k+\nu+1-N(N+\nu))/2}}{\Gamma((N(N+\nu)-2k-\nu-1)/2)}\times\\
\fl&&\times x^{k+(\nu-1)/2}
\left(1-Nx\right)^{(N(N+\nu)-2k-\nu-1)/2-1}\ \theta(1-Nx).\label{generalformula}
\end{eqnarray}
where $\theta(z)$ is the Heaviside step function. This equation is our first main result of this section.
The algorithm to compute $p_{N,\nu}^{(\mathrm{FT})}(x)$ for odd $\nu$ works as follows:
\begin{enumerate}
\item Compute the polynomial $h_{N,\nu}(x)$ for the sought $(N,\nu)$ using Edelman's recursion relation \cite{Edelrec};
\item Extract the coefficients $\mathfrak{q}_k$ of the polynomial from (\ref{hedel});
\item Insert these coefficients in the general formula (\ref{generalformula}).
\end{enumerate}

To give some explicit examples we have worked out in all detail the cases $\nu=1,3$, which are based on eq. (\ref{hnu13}):
\begin{eqnarray}
\fl p_{N,\nu=1}^{(\mathrm{FT})}(x) &=&\frac{N C_{N,N+1}^{(1)} 2^{-N(N+1)/2}(1-Nx)^{-2+N(N+1)/2}}{K_{N,N+1}^{(1)}\Gamma(-1+N(N+1)/2)}
\ \theta(1-Nx)\ ,\label{nu1p}\\
\fl p_{N,\nu=3}^{(\mathrm{FT})}(x) &=&\frac{C_{N,N+3}^{(1)}}{K_{N,N+3}^{(1)} }\frac{\Gamma(3+N)}{2(N+1)(N-1)!}\ \theta(1-Nx)\sum_{k=0}^{N-1}\frac{(1-N)_k (-1)^k}{\Gamma(4+k) k!}\times\nonumber\\
\fl&&\times \frac{2^{2+k-N(N+3)/2}}{\Gamma(-2-k+N(N+3)/2)}\ x^{1+k}(1-Nx)^{-3-k+N(N+3)/2}\ .
\label{nu3p}
\end{eqnarray}

We stress, however, that the general formula (\ref{generalformula}) and the algorithm above
provide an explicit and user-friendly solution to the problem of computing $p_{N,\nu}^{(\mathrm{FT})}(x)$ for {\em any} desired value of $(N,\nu)$. The obtained results are much more explicit and manageable than 
those expressed in terms of a finite sum over partitions in \cite{chen}. 
To illustrate our algorithm we have generated examples with higher $\nu$.

\begin{figure}[h]
  \centerline{
    \mbox{\includegraphics[scale=1.00]{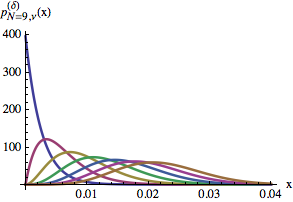}}
  }
  \caption{Smallest eigenvalue density $p_{N,\nu}^{(\mathrm{FT})}(x)$ for $N=9$ and   $\nu=1,3,5,\ldots,13$ odd, using formula (\ref{generalformula})
  and a simple Mathematica$\textsuperscript{\textregistered}$ implementation of our algorithm above. The running time for the full evaluation was less than $10$ sec. on a standard laptop.}
  \label{densitynuodd}
  \end{figure}

We wrote a simple Mathematica$\textsuperscript{\textregistered}$ code that generates
explicit expressions for the smallest eigenvalue density
(the code is freely available on demand).
A plot for different values of odd $\nu$ and $N=9$ is provided in fig. \ref{densitynuodd}.

\subsection{Even $\nu=M-N$}
\label{sectnueven}

Now, we turn to the more complicated case $\nu=M-N\geq0$ even. Also in this case, we can derive a general formula for $p_{N,\nu}^{(\mathrm{FT})}(x)$
involving the {\em same} polynomial coefficients as in Edelman's recursion.
To the best of our knowledge, apart from $\nu=0$ \cite{majbohi}, no such expressions were previously known for the \textsf{FTWL} ensemble.

Edelman \cite{Edelrec} states that the smallest eigenvalue distribution for a \textsf{WL} ensemble with $\beta=1$
and $\nu$ even can be written as follows:
\begin{equation}\label{pnuedel2}
\fl p_{N,\nu}^{(\mathrm{WL})}(x) = 2^{-\frac12}
c_{N,\nu} x^{(\nu-1)/2}e^{-Nx/2}[f_{N,\nu}(x)\mathcal{U}_1(x)+g_{N,\nu}(x)\mathcal{U}_2(x)]
\ ,
\end{equation}
where $f_{N,\nu}(x)$ and $g_{N,\nu}(x)$ are polynomials of degree at most $(N-1)\nu/2$, and $c_{N,\nu}$ is given in eq. (\ref{cNnu}). The two related functions $\mathcal{U}_1(x)$ and $\mathcal{U}_2(x)$
are defined as follows:
\begin{eqnarray}
\mathcal{U}_1(x) &\equiv\mathrm{U}\left(\frac{N-1}{2},-\frac{1}{2},\frac{x}{2}\right)
\ ,\non\\
\mathcal{U}_2(x) &\equiv -\ \frac{N-1}{4}\  \mathrm{U}\left(\frac{N+1}{2},\frac{1}{2},\frac{x}{2}\right)
=\ \mathcal{U}_1(x)^{\,\prime}\ ,
\label{Udef}
\end{eqnarray}
where $U(a,b,z)$ is a Tricomi confluent hypergeometric function given by:
\begin{equation}\label{tricomih}
\mathrm{U}(a,b,z)=\frac{1}{\Gamma(a)}\int_0^\infty d\tau e^{-z\tau}\tau^{a-1}(\tau+1)^{-a+b-1}\ .
\end{equation}

Let us now write the polynomials $f_{N,\nu}(x)$ and $g_{N,\nu}(x)$ as:
\begin{eqnarray}\label{fgedel}
f_{N,\nu}(x) &=\sum_{k=0}^{\frac{\nu}{2}(N-1)} \mathfrak{f}_k x^k\ ,\label{fgedel1}\\
g_{N,\nu}(x) &=\sum_{k=0}^{\frac{\nu}{2}(N-1)} \mathfrak{g}_k x^k\ ,\label{fgedel2}
\end{eqnarray}
with some rational coefficients $\mathfrak{f}_k$ and $\mathfrak{g}_k$, which depend also on $N,\nu$. As in the odd $\nu$ case they follow from a recurrence relation given in \cite{Edelrec} 
along with a Mathematica$\textsuperscript{\textregistered}$ code to generate them. The first two examples read:
\begin{eqnarray}
\fl f_{N,\nu=0}(x)&=&\frac{\Gamma\left(N\right)\sqrt{\pi}}{2^{N-1}\Gamma\left(\frac{N}{2}\right)\Gamma\left(\frac{N+1}{2}\right)}\ ,\ \ g_{N,\nu=0}(x)=0\ ,
\label{fgnu0}\\
\fl f_{N,\nu=2}(x)&=&\frac{2^{N}\Gamma\left(\frac{N+1}{2}\right)\Gamma\left(\frac{N+2}{2}\right)}{N\sqrt{\pi}}L_{N-1}^{(2)}(-x)\
\non , \\
\fl g_{N,\nu=2}(x)&=&-\ 
\frac{2^{N+1}\Gamma\left(\frac{N+1}{2}\right)\Gamma\left(\frac{N+2}{2}\right)}{N(N-1)\sqrt{\pi}}\ xL_{N-2}^{(3)}(-x)\ .
\label{fgnu2}
\end{eqnarray}

Applying now eq. (\ref{mainlapp}) to (\ref{pnuedel2}), we get:
\begin{equation}
\fl\mathcal{L}\left[{p}_{N,\nu}^{(\mathrm{FT})}(x;t)\right](s) =\frac{C_{N,M}^{(1)}}{K_{N,M}^{(1)}} 2^{-\frac12}c_{N,\nu}\sum_{k=0}^{\frac{\nu}{2}(N-1)} \left[\mathfrak{f}_k \mbox{\boldmath{$\Phi$}}_k(x,s)-\mathfrak{g}_k \mbox{\boldmath{$\Psi$}}_k(x,s)\right]\ ,
\end{equation}
where
\begin{eqnarray*}
\fl\mbox{\boldmath{$\Phi$}}_k(x,s) &\equiv& x^{k+(\nu-1)/2}e^{-sNx}(2s)^{k+1+\frac{\nu-1-N(N+\nu)}{2}}\mathcal{U}_1(2sx)\non\\
\fl&=&\frac{x^{k+(\nu-1)/2}}{\Gamma\left(\frac{N-1}{2}\right)}\int_0^\infty d\tau\ \tau^{\frac{N-3}{2}}(\tau+1)^{-\frac{N}{2}-1}
(2s)^{k+\frac{\nu+1-N(N+\nu)}{2}}e^{-sx(N+\tau)}
\ ,\\
\fl \mbox{\boldmath{$\Psi$}}_k(x,s) &\equiv& \frac{N-1}{4}\ x^{k+(\nu-1)/2}e^{-sNx}(2s)^{k+1+\frac{\nu-1-N(N+\nu)}{2}}\mathcal{U}_2(2sx)
\non\\
\fl&=&\frac{N-1}{4}\frac{x^{k+(\nu-1)/2}}{\Gamma\left(\frac{N+1}{2}\right)}\int_0^\infty d\tau\ \tau^{\frac{N-1}{2}}(\tau+1)^{-\frac{N}{2}-1}
(2s)^{k+\frac{\nu+1-N(N+\nu)}{2}}e^{-sx(N+\tau)}
.
\end{eqnarray*}
Here we have inserted the integral representation (\ref{tricomih}).

Taking the inverse Laplace transform of the quantities in square brackets, and setting $t=1$ we can eventually write:
\begin{equation}\label{eventual}
\fl p_{N,\nu}^{(\mathrm{FT})}(x)=\frac{C_{N,M}^{(1)}}{K_{N,M}^{(1)}} 2^{-\frac12}c_{N,\nu}\sum_{k=0}^{\frac{\nu}{2}(N-1)}
\frac{2^{\frac{\nu+1+2k-N(N+\nu)}{2}}}{\Gamma\left(\frac{N(N+\nu)-\nu-1-2k}{2}\right)}\left[\mathfrak{f}_k \mbox{\boldmath{$\Theta$}}_k(x)-\mathfrak{g}_k \mbox{\boldmath{$\Xi$}}_k(x)\right]\ ,
\end{equation}
where:
\begin{eqnarray}
\fl\mbox{\boldmath{$\Theta$}}_k(x) &\equiv&\frac{x^{k+(\nu-1)/2}}{\Gamma\left(\frac{N-1}{2}\right)}\int_0^\infty d\tau\ \frac{\tau^{\frac{N-3}{2}}}{(\tau+1)^{\frac{N}{2}+1}}
(1-x(N+\tau))^{\frac{N(N+\nu)-\nu-3-2k}{2}}\theta(1-x(N+\tau))
\non\\
\fl&=&
\frac{\Gamma\left(\frac{N(N+\nu)-\nu-1-2k}{2}\right)}{\Gamma
\left(\frac{N(N+\nu)+N-\nu-2-2k}{2}\right)}
\ x^{k+\frac{\nu-N}{2}}(1-Nx)^{\frac{N(N+\nu)-\nu+N-4-2k}{2}}\times
\non\\
\fl&&\times\ _2 F_1\left(\frac{N-1}{2},\frac{N}{2}+1;\frac{-2k+(N-1)(N+\nu+2)}{2};N-\frac{1}{x}\right)
\label{Theta}\\
\fl\mbox{\boldmath{$\Xi$}}_k(x) &\equiv&\frac{N-1}{4}\frac{x^{k+(\nu-1)/2}}{\Gamma\left(\frac{N+1}{2}\right)}\int_0^\infty d\tau\ \tau^{\frac{N-1}{2}}(\tau+1)^{-\frac{N}{2}-1}\times
\non\\
\fl&&\times
(1-x(N+\tau))^{\frac{N(N+\nu)-\nu-3-2k}{2}}\theta(1-x(N+\tau))
\non\\
\fl&=&\frac{N-1}{4}
\frac{\Gamma\left(\frac{N(N+\nu)-\nu-1-2k}{2}\right)}{\Gamma
\left(\frac{N(N+\nu)+N-\nu-2k}{2}\right)}
\ x^{k-1+\frac{\nu-N}{2}}(1-Nx)^{-k+\frac{(N-1)(N+\nu+2)}{2}}\times
\non\\
\fl&&\times\ _2 F_1\left(\frac{N+1}{2},\frac{N}{2}+1;\frac{-2k-\nu+N(1+N+\nu)}{2};N-\frac{1}{x}\right).
\label{Xi}
\end{eqnarray}
In the second step we have performed the integrals and applied some simple algebra,
and $_2 F_1(a,b;c;z)$ is a standard hypergeometric function.

Eq. (\ref{eventual}) 
together with eqs. (\ref{Theta}) and (\ref{Xi}) is the second result of this section.
Again,  the algorithm to compute $p_{N,\nu}^{(\mathrm{FT})}(x)$ for even $\nu$ works as follows:
\begin{enumerate}
\item Compute the polynomials $f_{N,\nu}(x)$ and $g_{N,\nu}(x)$ for the sought $(N,\nu)$ using Edelman's recursion \cite{Edelrec};
\item Extract the coefficients $\mathfrak{f}_k$ and $\mathfrak{g}_k$ of the polynomial from (\ref{fgedel1}) and (\ref{fgedel2});
\item Insert these coefficients in the general formula (\ref{eventual}).
\end{enumerate}

To provide explicit examples we give the expressions for $\nu=0,2$ based on eqs. (\ref{fgnu0})
and (\ref{fgnu2}), where the result for $\nu=0$ was previously derived in \cite{majbohi}:
\begin{eqnarray}
\fl p_{N,\nu=0}^{(\mathrm{FT})} (x) &=&\frac{N\Gamma(N) \Gamma(N^2/2)}{2^{N-1}\Gamma(N/2)\Gamma((N^2+N-2)/2)}
{x^{-N/2}} {(1-Nx)^{(N^2+N-4)/2}}
\times\non\\
\fl&&\times{_2F_1}\left(\frac{N+2}{2},\frac{N-1}{2};\frac{N^2+N-2}{2};N-\frac{1}{x}\right)
\label{nu0p}
\end{eqnarray}
and
\begin{eqnarray}
\fl p_{N,\nu=2}^{(\mathrm{FT})} (x) &=&
\frac{C_{N,N+2}^{(1)}}{K_{N,N+2}^{(1)}}\frac{\Gamma((N+1)/2)}{\sqrt{2\pi}}
\sqrt{x}\left(\mbox{\boldmath{$\phi$}}_N(x)+
\mbox{\boldmath{$\psi$}}_N(x)\right)\ ,\label{nu2p}
\end{eqnarray}
where:
\begin{eqnarray}
\fl\nonumber\mbox{\boldmath{$\phi$}}_N(x) &=&2^{(3-N(N+2))/2}x^{(1-N)/2}(1-Nx)^{-3+3N/2+N^2/2}
\sum_{k=0}^{N-1}\frac{2^k (N+1)!}{k!(N-1-k)!(k+2)!}
\times\\
\fl&&\times\left(\frac{x}{1-Nx}\right)^k
\ _2 F_1\left(\frac{N-1}{2},1+\frac{N}{2};\frac{N^2}{2}+\frac{3N}{2}-2-k;N-\frac{1}{x}\right)\ ,
\\
\fl\nonumber\mbox{\boldmath{$\psi$}}_N(x) &=&2^{(3-N(N+2))/2}x^{(1-N)/2}(1-Nx)^{-3+3N/2+N^2/2}\sum_{k=0}^{N-2}\frac{2^k (N+1)!}{k!(N-2-k)!(k+3)!}
\times\\
\fl&&\times\left(\frac{x}{1-Nx}\right)^k
\ _2 F_1\left(\frac{N+1}{2},1+\frac{N}{2};\frac{N^2}{2}+\frac{3N}{2}-2-k;N-\frac{1}{x}\right)\ .
\end{eqnarray}

\begin{figure}[h]
  \centerline{
    \mbox{\includegraphics[scale=0.90]{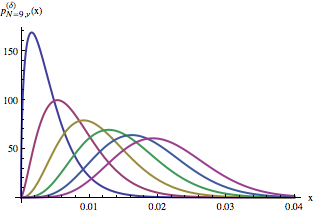}}
  }
  \caption{Smallest eigenvalue density $p_{N,\nu}^{(\mathrm{FT})}(x)$ for $N=9$ and $\nu=2,4,6,\ldots,12$ even using formula (\ref{eventual})
  and a simple Mathematica$\textsuperscript{\textregistered}$ implementation of the algorithm above. The running time for the full evaluation was less than $50$ sec. on a standard laptop.}
  \label{densitynueven}
  \end{figure}

In order to illustrate our algorithm above
we again wrote a simple Mathematica$\textsuperscript{\textregistered}$ code that generates
explicit expressions for the smallest eigenvalue density.
A plot for different even $\nu>0$ and $N=9$ as an example is provided in fig. \ref{densitynueven}.

\subsection{Moments for odd $\nu$}
\label{MomentsN}

Arbitrary moments $\langle \lambda_{\mathrm{min}}^\ell\rangle_{N,\nu,\beta=1}^{(\mathrm{FT})}$ can be computed easily in a closed form for odd $\nu$ from eq. (\ref{generalformula}), using the following integral formula:
\begin{equation}
\int_0^{1/N}dx\ x^{\omega-1}(1-Nx)^{\xi-1}=N^{-\omega}\mathrm{B}(\omega,\xi)\ .
\end{equation}
Here $\mathrm{B}(\omega,\xi)$ is Euler's Beta function.
We thus obtain
\begin{eqnarray}
\fl\nonumber
\langle \lambda_{\mathrm{min}}^\ell\rangle_{N,\nu=2m+1,\beta=1}^{(\mathrm{FT})}
&=&\frac{C_{N,M}^{(1)}}{K_{N,M}^{(1)}} 2^{\frac{N}{2}-1}c_{N,2m+1}
\sum_{k=0}^{(N-1)m} \frac{\mathfrak{q}_k\ 2^{m+1+(2k-N(N+\nu))/2}}{\Gamma(\frac{N(N+2m+1)}{2}-k-m-1)}\times \\
\fl&&\times
\mathrm{B}\left( \ell+k+1+m,\frac{N(N+2m+1)}{2}-k-m-1\right) .
\label{generalmoment}
\end{eqnarray}
To give an example, the $\ell$th moment for $\nu=1$ is given as follows:
\begin{eqnarray*}
\langle \lambda_{\mathrm{min}}^\ell\rangle_{N,\nu=1,\beta=1}^{(\mathrm{FT})}
=\frac{C_{N,N+1}^{(1)}}{K_{N,N+1}^{(1)}}2^{-N(N+1)/2}\frac{\Gamma(1+\ell)}{N^\ell\Gamma(\ell+ N(N+1)/2)}\ .
\end{eqnarray*}
In particular for the first moment $(\ell=1)$ or average value we get the following answer
\begin{equation}
\langle \lambda_{\mathrm{min}}\rangle_{N,\nu=1,\beta=1}^{(\mathrm{FT})}=\frac{2}{N^2(N+1)}\sim 2/N^3\qquad\mbox{for } N\gg 1 \ .
\end{equation}
For comparison at $\nu=0$ the large-$N$ behavior is as follows \cite{majbohi}:
\begin{equation}
\langle \lambda_{\mathrm{min}}\rangle_{N,\nu=0,\beta=1}^{(\mathrm{FT})}\sim \mathfrak{c}/N^3\qquad\mbox{for } N\gg 1 \ ,
\end{equation}
where:
\begin{equation}
\mathfrak{c}=2\left[1-\sqrt{\frac{\pi e}{2}}\mathrm{erfc}(1/\sqrt{2})\right]\approx 0.688641\ldots \ .
\end{equation}
(compare with subsection \ref{Momentsoo}). Although the computation of moments for even $\nu$ is possible, based on our explicit expressions given previously, 
we did not find
short closed expressions as in eq. (\ref{generalmoment}) above.

\subsection{Numerical checks}
\label{numcheck}

\begin{figure}[h]
  \centerline{
    \mbox{\includegraphics[scale=0.35]{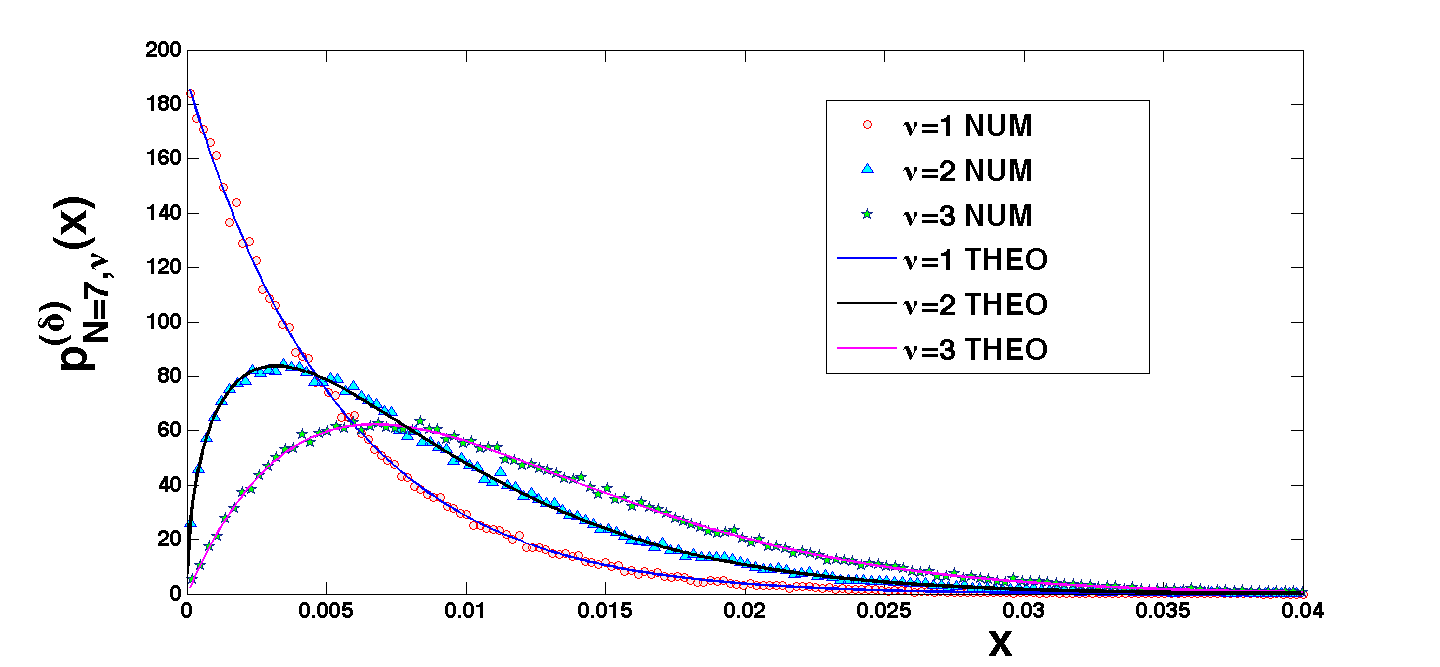}}
  }
  \caption{The density $p_{N=7,\nu}^{(\mathrm{FT})}(x)$ of the smallest Schmidt eigenvalue from numerical simulations compared with our predictions eqs. (\ref{nu1p}), (\ref{nu2p}), and (\ref{nu3p})
  for $\nu=1,2,3$ respectively (solid lines).}
  \label{density1}
  \end{figure}

In fig. \ref{density1} we compare a few theoretical densities from the previous subsections \ref{sectnuodd} and \ref{sectnueven} for finite $N$ with the corresponding numerical results. These are obtained as follows \cite{ZS,ZyckBook}:
\begin{enumerate}
\item we generate $n\approx 10^5$ {\em real} Gaussian $M\times N$ matrices $\mathcal{X}$ (where $N=7,M=N+\nu$).
\item for each instance we construct the
Wishart matrix $\mathcal{W}=\mathcal{X}^T \mathcal{X}$.
\item we diagonalize $\mathcal{W}$ and collect its $N$ real and non-negative eigenvalues $\{ \lambda_1,\ldots,\lambda_N\}$.
\item we define a new variable $0\leq \mu_1\leq 1$ as $\mu_1 =\lambda_{\mathrm{min}} /\sum_{i=1}^N \lambda_i$.
\item we construct a normalized histogram of $\mu_1$.
\end{enumerate}
The agreement between theory and simulations is excellent.

\section{Equivalence proofs and new results in the large-$N,M$ limit}
\label{equiv}

We now turn to the microscopic large-$N,M$ limit for the orthogonal and unitary ensembles ($\beta=1,2$),
with $\nu=M-N\geq 0$ fixed.  More precisely, let us define the following microscopic scaling limits at the origin:
\begin{eqnarray}
\fl Q_\nu^{(\mathrm{FT})}(y) &=&\lim_{N,M\to\infty} q_{N,\nu}^{(\mathrm{FT})}\left(\frac{y}{4N^3}\right)
\ ,\ \
P_\nu^{(\mathrm{FT})}(y) =\lim_{N,M\to\infty} \frac{1}{4N^3}
p_{N,\nu}^{(\mathrm{FT})}\left(\frac{y}{4N^3}\right)
\label{aspp}\\
\fl Q_\nu^{(\mathrm{WL})}(y) &=&\lim_{N,M\to\infty} q_{N,\nu}^{(\mathrm{WL})}\left(\frac{y}{4N}\right)\ ,\ \
P_\nu^{(\mathrm{WL})}(y) =\lim_{N,M\to\infty} \frac{1}{4N}p_{N,\nu}^{(\mathrm{WL})}\left(\frac{y}{4N}\right)
\label{WLlim}
\end{eqnarray}
Notice that the mean level spacing at the origin (the $N$-dependent rescaling factor) is 
different for \textsf{FTWL} and standard \textsf{WL} ensembles. This fact was also observed when comparing the corresponding macroscopic 
densities in the Gaussian ensembles with and without constraint \cite{ACMV}.

It has been shown that in the microscopic limit (i) the smallest eigenvalue distribution in non-Gaussian generalizations of \textsf{WL} ensembles is universal 
\cite{DNW,DN98}, and (ii) for all real $\beta>0$, the microscopic limit 
for the smallest eigenvalue distribution is the same for \textsf{WL} and \textsf{FTWL} ensembles in all cases where a representation in terms of \textsf{HFMA} exists \cite{chen}.
We also recall that universality (ii), that is the independence of the constraint, was shown earlier for all microscopic
density correlation functions in the bulk of the spectrum for unitary non-chiral ensembles
for monic even potentials \cite{AV} 
or i.i.d. matrix elements \cite{Goetze,Liu}.

Building on \cite{chen}, from now on we shall assume that:
\begin{eqnarray}
 Q_\nu^{(\mathrm{WL})}(y) &=& Q_\nu^{(\mathrm{FT})}(y)\, \\
 P_\nu^{(\mathrm{WL})}(y) &=& P_\nu^{(\mathrm{FT})}(y)\ ,
\label{univrel}
\end{eqnarray}
for all $\beta$ and $y$. Therefore, there is no reason to keep the superscripts $^{(\mathrm{WL})}$ and $^{(\mathrm{FT})}$. We will thus use the unified notation $Q_\nu^{(\beta)}(y)$ and $P_\nu^{(\beta)}(y)$, labeled by $\beta$ only\footnote{
We note in passing that the large $y\gg 1$ behavior of $Q_\nu^{(\beta)}(y)$ is known
from \cite{tracywidom}, see also \cite{chenmanning} for the calculation of leading term using a Coulomb gas method.}.
Note that one still has
\begin{eqnarray}
\label{derivppp} P_\nu^{(\beta)}(y)=-Q_\nu^{(\beta)}(y)^\prime\ ,\\
\int_0^\infty dy P_\nu^{(\beta)}(y)=1\ ,\\
Q_\nu^{(\beta)}(0) = 1\ .
\label{anti}
\end{eqnarray}
What are the known results for $Q_\nu^{(\beta)}(y)$ and $P_\nu^{(\beta)}(y)$? It turns out that two communities have derived different formulae for these very same quantities using different languages: 
\textsf{HFMA} on the one side, and determinants or Pfaffians of Bessel functions on the other side (see two next subsections).
It is one of the goals of this paper to highlight this connection, which is not widely appreciated, and to actually prove the equivalence of these formulae by simple algebraic methods. 
As a result of this (and of the previous section) we will derive some new compact results in the second language in cases where only \textsf{HFMA} or no results were previously known.

We should also mention here that a third way exists to compute the cumulative distribution. Formally it is given 
by the Fredholm determinant of the corresponding Bessel kernel. 
This has been mapped to expressions containing a transcendent of the solution of the Painlev\'e V 
for $\beta=2$ \cite{tracywidom} and $\beta=1,4$ \cite{forr:pfaff}. In particular, the distributions 
for different $\beta$s can be related to each other. However, this relation does not allow us to generate more explicit expressions 
in terms of determinants and Pfaffians below.

\subsection{Hypergeometric functions of matrix argument (\textsf{HFMA})}
\label{sectHFMA}

In order to establish the claimed equivalence we first need to state the results in the first language.
Forrester \cite{forrwaive} and Chen {\em et al.} \cite{chen} independently gave expressions for $P_\nu^{(\beta)}(y)$ and $Q_\nu^{(\beta)}(y)$ valid for all $\beta>0$ and integer $m=(\beta/2)(\nu+1)-1\in\mathbb{N}$, in terms of \textsf{HFMA} as:
\begin{eqnarray}
\label{nscaled1} Q_\nu^{(\beta)}(y) &=e^{-\beta y/8}\ _0 F_1^{(\beta/2)}(-;2m/\beta;(y/4)\mathbb{I}_m)\ ,\\
P_\nu^{(\beta)}(y) &=A_{m,\beta}\ y^m e^{-\beta y/8}\ _0 F_1^{(\beta/2)}(-;2+2m/\beta;(y/4)\mathbb{I}_m)\ ,
\label{nscaledP}
\end{eqnarray}
where the constant $A_{m,\beta}$ is given by\footnote{Note that the expression for the constant $A_{m,\beta}$, eq. (2.15c) in \cite{forrwaive} (and subsequently in \cite{chen}), as well
as eq. (\ref{Qybessel1}) as it appears in \cite{chen} contain misprints.}:
\begin{equation}
A_{m,\beta} =
\frac{1}{4^{m+1}} (\beta/2)^{2m+1}\frac{\Gamma(\beta/2+1)}{\Gamma(m+1)\Gamma(m+1+\beta/2)}\ ,
\label{Adef}
\end{equation}
and $\mathbb{I}_m$ is the identity matrix $m\times m$. For definitions and properties of \textsf{HFMA} we refer to Appendix \ref{hypA}. We note that for $\beta=1$ integer $m$ implies that $\nu=M-N$ must be odd.

The formulas above were specialized in \cite{chen} to the two cases (a) $m=1$ for  all $\beta>0$, and (b) for $\beta=\nu=2$,  yielding respectively:
\begin{eqnarray}
\mbox{(a)}\ \ Q_{\nu=(4/\beta)-1}^{(\beta)}(y) &=2^{-1+2/\beta}\Gamma(2/\beta)e^{-\beta y/8}y^{1/2-1/\beta}\mathrm{I}_{\frac{2}{\beta}-1}(\sqrt{y})\label{Qybessel1}\\
\mbox{(b)}\ \ Q_{\nu=2}^{(\beta=2)}(y) &=e^{-y/4}\left(\mathrm{I}_0(\sqrt{y})^2-\mathrm{I}_1(\sqrt{y})^2\right)\label{Qybessel2}
\end{eqnarray}
where $\mathrm{I}_\rho(x)$ is a modified Bessel function of the first kind. The appearance of Bessel functions for special instances of formulas (\ref{nscaled1}) and (\ref{nscaledP}) is not at all accidental, as we will see now.

\subsection{Determinants and Pfaffians of Bessel functions.\label{besselsec}}

In the context of Random Matrix Theory applied to effective theories of Quantum Chromodynamics (\textsf{QCD}) or \textsf{QCD}-like theories, the large-$N$ distribution of the smallest Dirac operator eigenvalue 
has been studied intensely for the symmetry classes $\beta=1,2$ (and 4), 
see \cite{Jac} for references. It turns out that the Dirac eigenvalues $y_i$ (occurring in $\pm$ pairs due to chiral symmetry) are precisely distributed according to the \textsf{WL} \textsf{jpdf}
(\ref{wishart11}) upon the mapping $y_i^2=\la_i$.
The relation between the smallest Dirac eigenvalue $\mathfrak{P}_\nu^{(\beta)}(s)$, and the smallest \textsf{WL} eigenvalue distribution trivially follows:
\begin{eqnarray}\label{diractowl}
\mathfrak{P}_\nu^{(\beta)}(s)&=&2s P_\nu^{(\beta)} (s^2)\\
\mathfrak{Q}_\nu^{(\beta)}(s)&=&Q_\nu^{(\beta)} (s^2) \ ,
\label{diractowl2}
\end{eqnarray}
where we also give the relation to the cumulative distribution $\mathfrak{Q}_\nu^{(\beta)}(s)$ in the Dirac picture. We still have $\mathfrak{P}_\nu^{(\beta)}(s)=-\mathfrak{Q}_\nu^{(\beta)\prime}(s)$
and the standard normalization $\int_0^\infty ds\ \mathfrak{P}_\nu^{(\beta)}(s)=1$ is ensured.

The following results hold for our symmetry classes:
\begin{eqnarray}
\fl\mathfrak{P}_{\nu}^{(\beta=2)}(s)&=& \,\frac{s}{2}\ e^{-s^2/4}\det
\left[\mathrm{I}_{i-j+2}(s)\right]_{i,j=1,\ldots,\nu}\ ,\ \ \nu\in\mathbb{N}\ ,
\label{pb2}\\
\fl\mathfrak{P}_{\nu=2m+1}^{(\beta=1)}(s) &=& C_m s^{1-m} e^{-s^2/8}
\mathrm{Pf}\left[ (i-j)\mathrm{I}_{i+j+3}(s)\right]_{i,j=-m+\frac12,\ldots, m-\frac12}
\ ,\ m\in\mathbb{N},
\label{pb3}\\
\fl\mathfrak{P}_{\nu=0}^{(\beta=1)}(s)&=& \frac14(2 + s)e^{-s^2/8-s/2}\ ,
\label{pb1}\\
\fl\mathfrak{P}_{\nu=2}^{(\beta=1)}(s)&=& \frac14\Big((2 +
  s)\mathrm{I}_2(s)+s\mathrm{I}_3(s) \Big)e^{-s^2/8-s/2}\ ,
\label{pbus}
\end{eqnarray}
where $\mathrm{Pf}(\mathcal{A})=\sqrt{\det\mathcal{A}}$ stands for the Pfaffian of the antisymmetric matrix $\mathcal{A}$.

Most of these results were known in the literature, apart from the normalization constant $C_m$ given in eq. (\ref{Cm}), and the new eq. (\ref{pbus}) to be derived below.
For $\beta=2$ the distribution eq. (\ref{pb2}) was derived in
\cite{forrhughes,WGW,DNW}.
For $\beta=1$ and $\nu=0$ eq. (\ref{pb1}) was shown in \cite{Forrester1st},
and then extended to all odd $\nu$ in \cite{DN98} up to normalization.

We note in passing that for $\beta=4$ the distribution for $\nu=0$ is known
explicitly from \cite{Forrester1st} (see eq. (\ref{Qybessel1}) and take its derivative), but the results quoted in \cite{nagaoforr}
for $\beta=4$
only hold for $m=\beta(\nu+1)/2-1$ even, that is for half-integer values of $\nu$.
Finally we mention that for small $\nu$ all distributions for $\beta=1,2,4$ can be very well approximated by using finite $N=2$ results \cite{ABPS}, like in the Wigner surmise.

Turning to the corresponding cumulative distributions the following results hold:
\be
\mathfrak{Q}_\nu^{(\beta=2)}(s)= \ e^{-s^2/4}\det
\left[\mathrm{I}_{i-j}(s)\right]_{i,j=1,\ldots,\nu}\ ,\ \ \nu\in\mathbb{N}
\label{Qb2}
\ee
for $\beta=2$ \cite{DNW}, and
for $\beta=1$ and odd $\nu=2m+1$ our new result reads
\be
\fl\mathfrak{Q}_{\nu=2m+1}^{(\beta=1)}(s)=
2^m s^{-m}
e^{-s^2/8} \mathrm{Pf}\left[(j-k)\mathrm{I}_{1+k+j}(s)\right]_{j,k=-m+\frac12,\ldots, m-\frac12}
\ , m\in\mathbb{N}.
\label{Qb1}
\ee

Before providing a general algebraic proof of the equivalence of the two languages in subsections \ref{beta2} and \ref{beta1}, we can quickly check the agreement between the two different
formulations above, 
starting from the special \textsf{HFMA} cases in eqs. (\ref{Qybessel1}) and (\ref{Qybessel2}).
Using the map (\ref{diractowl2}),
eq. (\ref{Qb2}) obviously contains eq. (\ref{Qybessel2}) as a special case.
Specifying $\beta$ in eq. (\ref{Qybessel1}) leads to the special cases ($\beta=2,\nu=1$) and ($\beta=1,\nu=3$) in eqs. (\ref{Qb2}) and (\ref{Qb1}) respectively.

Additionally one can differentiate these and arrive at corresponding special cases in
eqs. (\ref{pb2}) and (\ref{pb1}) respectively, after using some Bessel identities.

\subsection{Results for the scaled
smallest eigenvalue distribution for $\beta=1$ and $\nu=0,2$.}\label{scaledsubnu}

In this subsection, we give two results for the scaled
smallest eigenvalue distribution at even $\nu=0$ and $\nu=2$. In principle the
universality eq. (\ref{univrel}), that is the agreement between the \textsf{WL}
and \textsf{FTWL} distributions in the microscopic large-$N$ limit, has not been shown for $\beta=1$ with even $\nu$,
although there is little doubt that it extends from the odd $\nu$ case. We
explicitly check here that this is indeed the case by taking the microscopic
large-$N$ limit for \textsf{FTWL} at $\nu=0$ starting from the known
eq. (\ref{nu0p}) for finite-$N$ and getting to the same result as in \textsf{WL}. For
$\nu=2$ we derive a new result that was unknown even for the microscopic
limit of the unconstrained \textsf{WL} ensemble.

For $\nu=0$ we first recall how the known microscopic result for \textsf{WL} is obtained. Starting from the finite-$N$ expression eq. (\ref{pnuedel2}) together with eq. (\ref{fgnu0}) 
we need to know the asymptotic limit of the Tricomi
confluent hypergeometric function, in the scaling limit eq. (\ref{WLlim}). It was derived in Corollary 3.1 of \cite{Edel}, alternatively it follows from eq. 13.3.3 \cite{Abramo}:
\be
\lim_{n\to\infty}\frac{2}{\sqrt{\pi}}\Gamma\left( \frac{n}{2}+1\right)
\mathrm{U}\left( \frac{n-1}{2},-\frac12,\frac{x}{2n}\right)=\left(1+\sqrt{x} \right)e^{-\sqrt{x}}\ .
\label{Ulim}
\ee
Collecting all prefactors we thus obtain
\be
\lim_{N\to\infty}\frac{1}{4N}p_{N,\nu}^{(\mathrm{WL})}(y/4N)=\frac18\left(1+\frac{2}{\sqrt{y}}\right)
e^{-y/8-\sqrt{y}/2}\ .
\label{WLb1nu0lim}
\ee
The relation eq. (\ref{diractowl}) maps this to the eq. (\ref{pb1}).
This result has to be compared with the limit eq. (\ref{aspp}) of eq. (\ref{nu0p}) in \textsf{FTWL} ensemble. Because the limit of the hypergeometric function is more involved we defer the derivation to Appendix \ref{nu0lim}, finding complete agreement as in eq. (\ref{univrel}):
\begin{eqnarray}
\fl \non P_{\nu=0}^{(\beta=1)}(y)\equiv 
P_{\nu=0}^{(\mathrm{FT})}(y) =\lim_{N\to\infty}\frac{1}{4N^3}p_{N,\nu}^{(\mathrm{FT})}\left(\frac{y}{4N^3}\right)=
P_{\nu=0}^{(\mathrm{WL})}(y) =\lim_{N\to\infty}\frac{1}{4N}p_{N,\nu}^{(\mathrm{WL})}\left(\frac{y}{4N}\right)\\
\label{nu0univ}
\end{eqnarray}
This extends the expected universality to the case of even $\nu=0$.

For $\nu=2$ the corresponding microscopic limit of (\ref{nu2p}) for \textsf{FTWL} is already rather cumbersome, 
involving the asymptotic of a sum of
hypergeometric functions. We therefore restrict ourselves to compute the
corresponding microscopic limit in the \textsf{WL} ensemble - which is also new - and
conjecture that the universal relation eq. (\ref{nu0univ}) extends to $\nu=2$
and in fact to all higher even $\nu$'s. Combining eqs. (\ref{pnuedel2}) and
(\ref{fgnu2}) we have to compute
\begin{eqnarray}
\nonumber\fl &&\lim_{N\to\infty} \frac{1}{4N}p_{N,\nu=2}^{(\mathrm{WL})}\left(\frac{y}{4N}\right) =
\lim_{N\to\infty}
\frac{\Gamma\left(\frac{N+1}{2}\right)}{\sqrt{2\pi}}
\left(\frac{y}{4N}\right)^{\frac12}e^{-y/8}\times\\
\nonumber\fl &&\left[L_{N-1}^{(2)}\left(\frac{-y}{4N}\right)
\mathrm{U}\left(\frac{N-1}{2},-\frac{1}{2},\frac{y}{8N}\right)
\!\!+\!\!\ \frac{y}{8N}L_{N-2}^{(3)}\left(\frac{-y}{4N}\right)\frac{N-1}{4} \mathrm{U}\left(\frac{N+1}{2},\frac{1}{2},\frac{y}{8N}\right)
\right].
\label{PWLnu2lim}
\end{eqnarray}
Eq. (\ref{Ulim}) together with the Laguerre asymptotic for negative argument,
\be
\lim_{m\to\infty}m^{-a}L_m^{(a)}(-x/m)=x^{-a/2}\mathrm{I}_a(2\sqrt{x})\ ,
\ee
yields the following final answer,
\be
\fl P_{\nu=2}^{(\beta=1)}(y)\ \equiv P_{\nu=2}^{(\mathrm{FT})}(y)=P_{\nu=2}^{(\mathrm{WL})}(y)\ = \frac18\Big( (1+2/\sqrt{y})\mathrm{I}_2(\sqrt{y})+\mathrm{I}_3(\sqrt{y})\Big)
e^{-y/8-\sqrt{y}/2}\ .
\label{pwlnu2}
\ee
Using eq. (\ref{diractowl}) this is mapped to eq. (\ref{pbus}) as claimed.

\subsection{Equivalence proofs for $\beta=2$}
\label{beta2}

In this subsection and in the following, we provide an algebraic link between the \textsf{HFMA} and the Bessel determinant languages in the spirit of earlier works \cite{forr:pfaff,gupta}.
We start from the formula in eq. (\ref{nscaledP}) which expresses the scaled smallest eigenvalue distribution (with or without fixed trace constraint) in terms of \textsf{HFMA}:
\begin{equation}
P_{\nu\equiv m}^{(2)}(y) =A_{m, 2}\ y^m e^{-y/4}\ _0 F_1^{(1)}\left(-;m+2;\frac{y}{4}\mathbb{I}_m\right)\label{appc1}
\end{equation}
Next, we use the following integral representation due to Forrester \cite{forrwaive} valid for integer values of $\lambda$ and $c$:
\begin{eqnarray}
\fl _0 F_1^{(2/\lambda)} \left(-;c+\frac{\lambda}{2}(m-1);x\mathbb{I}_m\right)&=&
\hat{B}_m(c,\lambda)\left(\frac{1}{x}\right)^{(c-1)m/2}\left(\frac{1}{2\pi}\right)^m\times\non
\\
\fl &\times&\int_{[-\pi,\pi]^m}\prod_{j=1}^m d\theta_j e^{2\sqrt{x}\cos\theta_j}e^{\mathrm{i}(c-1)\theta_j}
\left|\Delta(e^{\mathrm{i}\vec{\theta}})\right|^\lambda, \label{cc2}
\end{eqnarray}
where we have defined
\begin{equation}
\hat{B}_m(c,\lambda)=\prod_{j=1}^m\frac{\Gamma(1+\lambda/2)\Gamma(c+\lambda (j-1)/2)}{\Gamma(1+\lambda j/2)}\ ,
\end{equation}
and the Vandermonde determinant of the angles
\begin{equation}
\Delta(e^{\mathrm{i}\vec{\theta}})\equiv\prod_{j<k}(e^{\mathrm{i}\theta_k}-e^{\mathrm{i}\theta_j})\ .
\end{equation}
Comparison between (\ref{cc2}) and (\ref{appc1}) yields $c=3$ and $\lambda=2$.
Denoting by  $^\star$ complex conjugation, we can use the Andr\'eief identity \cite{and}
\begin{equation}
\fl\int \prod_{j=1}^m d\theta_j \omega(\theta_j)\Delta(e^{\mathrm{i}\vec{\theta}})\Delta^\star(e^{\mathrm{i}\vec{\theta}})=m!\ \det\left[\int d\theta e^{\mathrm{i}\theta (\ell-k)}\omega(\theta)\right]_{\ell,k=1,\ldots,m}\ .
\end{equation}
We thus obtain for  the $m$-fold integral in the second line of eq. (\ref{cc2})
\begin{eqnarray}
\fl m!\ \det\left[\int_{-\pi}^\pi d\theta e^{\mathrm{i}\theta (\ell-k+2)}e^{2\sqrt{x}\cos\theta}\right]_{\ell,k=1,\ldots,m}=m!\ (2\pi)^m\ \det\left[\mathrm{I}_{\ell-k+2}(2\sqrt{x})\right]_{\ell,k=1,\ldots,m}\non\\
\fl&&\label{detIb}
\end{eqnarray}
where we have used the following integral representation for the modified Bessel function (valid for integer index only):
\begin{equation}
\mathrm{I}_n (z)=\frac{1}{\pi}\int_0^\pi dt\ e^{z\cos t}\cos(n t)
\ ,\ \ n\in\mathbb{N}\ .\label{besselint}
\end{equation}
Setting $x=y/4$ in (\ref{detIb}) and simplifying all prefactors, we eventually get from (\ref{appc1}):
\begin{equation}\label{ppp1}
P_{\nu\equiv m}^{(2)}(y) =\frac{1}{4}e^{-y/4}\det\left[\mathrm{I}_{i-j+2}(\sqrt{y})\right]_{i,j=1,\ldots,\nu}
,\ \nu\in\mathbb{N}\ .
\end{equation}
This is identical to eq. (\ref{pb2}) after switching to the Dirac picture eq. (\ref{diractowl}).

The proof for the cumulative distribution goes along the same lines, merely changing the coefficient in front and the index shift of the Bessel function. 
We obtain the following from
eq. (\ref{nscaled1}) by identifying $c=1$ and $\lambda=2$ in the representation eq. (\ref{cc2}) and
collecting all prefactors:
\begin{eqnarray}
\fl Q_\nu^{(\beta=2)}(y) &=e^{-y/4}\ _0 F_1^{(1)}(-;m;(y/4)\mathbb{I}_m)
\ =\ e^{-y/4}\det
\left[\mathrm{I}_{i-j}(\sqrt{y})\right]_{i,j=1,\ldots,\nu}\ ,\nu\in\mathbb{N}\label{Qafterref}
\end{eqnarray}
This corresponds to eq. (\ref{Qb2}) using the map eq. (\ref{diractowl2}). Note that it is highly non-trivial to derive eq. (\ref{Qafterref})
from eq. (\ref{ppp1}) using eq. (\ref{derivppp}).

\subsection{Equivalence proof and new results for $\beta=1$}
\label{beta1}

We start again from the formula in eq. (\ref{nscaledP})
expressing the scaled smallest eigenvalue distribution (with or without fixed trace constraint) in terms of \textsf{HFMA}:
\begin{equation}
\fl P_{\nu\equiv 2m+1}^{(\beta=1)}(y) =A_{m, 1}\ y^m e^{-y/8}\ _0 F_1^{(1/2)}\left(-;2m+2;\frac{y}{4}\mathbb{I}_m\right)\ ,\ \ m\in\mathbb{N}\ .\label{appc2}
\end{equation}
Next, we use again the integral representation in eq. (\ref{cc2}) and $x=y/4$ to write:
\begin{eqnarray}
\fl _0 F_1^{(1/2)} \left(-;2+2m;x\mathbb{I}_m\right)&=&
\hat{B}_m(4,4)\left(\frac{1}{x}\right)^{3m/2}\left(\frac{1}{2\pi}\right)^m\times\non\\
&\times&\int_{[-\pi,\pi]^m}\prod_{j=1}^m d\theta_j e^{2\sqrt{x}\cos\theta_j}e^{3\mathrm{i}\theta_j}
|\Delta(e^{\mathrm{i}\vec{\theta}})|^4\ .
\label{cc22}
\end{eqnarray}
Comparison between (\ref{cc2}) and (\ref{appc2}) indeed yields $c=4$ and $\lambda=4$. Next, we use  section 11.5 in \cite{Mehta}, combining eqs. (11.5.2) and (11.5.4) there to write
\begin{equation}
\prod_{j<k}|e^{\mathrm{i}\theta_k}-e^{\mathrm{i}\theta_j}|^4 = \det\left[e^{\mathrm{i}p\theta_j}\ p e^{\mathrm{i}p\theta_j}\right],
\end{equation}
where the determinant on the right hand side is of size $2m\times 2m$ and indices run as follows, $1\leq j\leq m$ and $p=-(m-1/2),-(m-3/2),\ldots, (m-3/2),m-1/2$.

Now we can define the two sets of functions $\{\phi_k(\theta)\}=e^{\mathrm{i}k\theta}$ and $\{\psi_k(\theta)\}=ke^{\mathrm{i}k\theta}$
and thus use the de Bruijn identity \cite{deB55}:
\begin{eqnarray}
&&\int \prod_{j=1}^m d x_j \omega(x_j)\det\left[\phi_i(x_j) \quad\psi_i(x_j)\right]_{1\leq i\leq 2m,1\leq j\leq m} \non \\
&&=m!\ \mathrm{Pf}\left[\int dx\ \omega(x) \left(\phi_i(x)\psi_j(x)-\phi_j(x)\psi_i(x)\right)\right]_{1\leq i,j\leq 2m}\ ,
\end{eqnarray}
to evaluate the $m$-fold integral in the second line of eq. (\ref{cc22}):
\begin{eqnarray}
\fl
&& m!\ \mathrm{Pf}\left[\int_{-\pi}^\pi d\theta\ e^{2\sqrt{x}\cos\theta}e^{3\mathrm{i}\theta} (e^{\mathrm{i}k\theta}j e^{\mathrm{i}j\theta}-e^{\mathrm{i}j\theta}k e^{\mathrm{i}k\theta})\right]\non\\
&&= m! (2\pi)^m \ \mathrm{Pf}\left[(j-k)\mathrm{I}_{3+k+j}(2\sqrt{x})\right]\label{pfIb}
\end{eqnarray}
where the indices $(k,j)$ all run from $-(m-1/2),-(m-3/2),\ldots, (m-3/2),(m-1/2)$, and we have used again the integral representation for the Bessel function (\ref{besselint}).

Setting $x=y/4$ in (\ref{pfIb}) and simplifying all prefactors, we eventually get for eq.  (\ref{appc2}):
\begin{equation}
\label{Pb1final}
P_{\nu=2m+1}^{(\beta=1)}(y) =C_m y^{-m/2} e^{-y/8} \mathrm{Pf}\left[(j-k)\mathrm{I}_{3+k+j}(\sqrt{y})\right]
\end{equation}
where $C_m$ is the normalization constant:
\begin{equation}
\label{Cm}
C_m=\frac{\sqrt{\pi} (2m+1)!!}{16\ \Gamma(m+3/2)}\ .
\end{equation}
Applying now eq (\ref{diractowl}), we get the distribution in the Dirac picture (\ref{pb3})
including its explicit normalization constant for any  $\nu$, which
was previously unavailable.

The proof for the cumulative distribution easily follows from (\ref{nscaled1}), and we only quote the result which is new,
after having identified $c=2$ and $\la=4$ in eq. (\ref{cc22}) instead:
\begin{eqnarray}
Q_{\nu=2m+1}^{(\beta=1)}(y) &=&e^{-y/8}\ _0 F_1^{(1/2)}(-;2m;(y/4)\mathbb{I}_m)
\non\\
&=&
2^m y^{-m/2}
e^{-y/8} \mathrm{Pf}\left[(j-k)\mathrm{I}_{1+k+j}(\sqrt{y})\right]\ .\label{Q1dop}
\end{eqnarray}
The range of indices is the same as in eq. (\ref{pfIb}). Once again, deriving eq. (\ref{Q1dop}) directly from eq. (\ref{Pb1final})
using eq. (\ref{derivppp}) is highly non-trivial.
\\

\subsection{Moments for large $N$}
\label{Momentsoo}

We point out a universal expression for the large $N$ decay of the $\ell$th
moment of the smallest eigenvalue, valid for both $\beta=1$ and 2:
\begin{equation}\label{largeNbeh}
\fl\langle\lambda_{\mathrm{min}}^\ell\rangle_{N,\nu,\beta}^{(\mathrm{FT})}
\sim\frac{\kappa_{\ell,\nu,\beta}}{(4N^{3})^{\ell}}\ ,\
\langle\lambda_{\mathrm{min}}^\ell\rangle_{N,\nu,\beta}^{(\mathrm{WL})}
\sim\frac{\kappa_{\ell,\nu,\beta}}{(4N)^{\ell}}
\qquad\mbox{for }N\to\infty
\end{equation}
where we have defined the universal coefficient
\begin{equation}
\label{kappadef}
\kappa_{\ell,\nu,\beta}=\int_0^\infty ds\ s^{2\ell} \mathfrak{P}_\nu^{(\beta)}(s)\ .
\end{equation}
The scaling with different powers of $N$ trivially follows from the different spacings in the definition of the microscopic limit eqs. (\ref{aspp}) and (\ref{WLlim}), as we will show now.
Starting from the definition of the average of the smallest eigenvalue for \textsf{FTWL} (\ref{momentdef}) we have
\begin{eqnarray}
\fl\langle \lambda_{\mathrm{min}}^\ell\rangle_{N,\nu,\beta}^{(\mathrm{FT})}=\int_0^{1/N} dx\ x^\ell p_{N,\nu}^{(\mathrm{FT})}(x)
=\frac{1}{(4N^3)^\ell}\int_0^{4N^2} dy\ y^\ell \ \frac{1}{4N^3} p_{N,\nu}^{(\mathrm{FT})}
\left(\frac{y}{4N^3}\right)
\end{eqnarray}
after making a change of variable $x=y/4N^3$. This implies, using the limit (\ref{aspp}):
\begin{equation}
\lim_{N\to\infty}\left[(4N^3)^{\ell}\langle \lambda_{\mathrm{min}}^\ell\rangle_{N,\nu,\beta}^{(\mathrm{FT})}\right]=\int_0^\infty dy\ y^\ell P_\nu^{(\beta)}(y)\ .
\end{equation}
The same argument for \textsf{WL} with a different change of variables, $x=y/4N$, leads to
\begin{equation}
\lim_{N\to\infty}\left[(4N)^{\ell}\langle \lambda_{\mathrm{min}}^\ell\rangle_{N,\nu,\beta}^{(\mathrm{WL})}\right]=\int_0^\infty dy\ y^\ell P_\nu^{(\beta)}(y)\ .
\end{equation}
Alternatively the universal right hand side can be expressed in the Dirac picture, using the map eq. (\ref{diractowl}), as it is given in eq. (\ref{kappadef}).

  \begin{figure}[h]
  \centerline{
    \mbox{\includegraphics[scale=0.40]{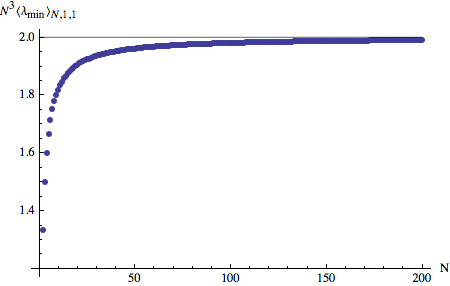}
    \includegraphics[scale=0.40]{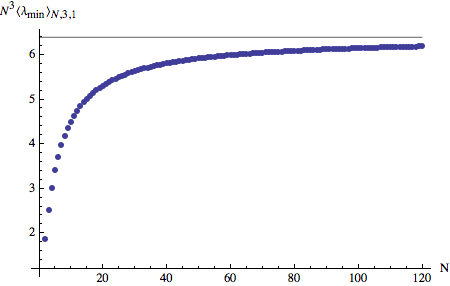}}
  }
  \caption{The behavior of $N^3 \langle \lambda_{\mathrm{min}}\rangle_{N,\nu,1}^{(\mathrm{FT})}$ for increasing $N$ and for $\nu=1,3$ (left and right panel respectively).
  The curves converge for $N\to\infty$ to the limiting values $(1/4)\
  \kappa_{1,1,1}=2$ (left) and $(1/4)\ \kappa_{1,3,1}=e^2-1\approx 6.38906$ (right).}
  \label{momentexpl}
  \end{figure}
Let us give a few examples, by simply inserting eqs. (\ref{pb1}) and (\ref{pb2}) into the integral
\begin{eqnarray}
\fl\nonumber\kappa_{\ell,\nu=0,\beta=1} &=&\frac{1}{4}\int_0^\infty ds \ s^{2\ell}(2+s)e^{-s^2/8-s/2}
\\
\fl&=&2^{3\ell} \left[-\sqrt{2}\ \ell \ \Gamma\left(1/2 + \ell\right)
\ _1 F_1\left(1/2 + \ell; 3/2; 1/2\right) \right.
\non\\
\fl&&\left.+
   \Gamma\left(1 + \ell\right) \Big( {_1 F_1}\left(1 + \ell; 1/2; 1/2\right) -
      \ _1F_1\left(1 + \ell; 3/2; 1/2\right)\Big)\right]
\\
\fl\kappa_{\ell,\nu=0,\beta=2} &=&\frac12\int_0^\infty ds \ s^{2\ell+1}e^{-s^2/4}= 4^\ell\Gamma(1+\ell)\ .
\end{eqnarray}
Specializing these results to the case of the first moment $\ell=1$ (average value), we obtain:
\begin{eqnarray}
\frac{1}{4}\ \kappa_{\ell=1,\nu=0,\beta=1} &= 2 - \sqrt{2 e \pi}\ \mathrm{erfc}\left(1/\sqrt{2}\right)\approx 0.688641...\\
\frac{1}{4}\ \kappa_{\ell=1,\nu=0,\beta=2} &=1\ ,
\end{eqnarray}
in complete agreement with \cite{majreview,majbohi} (after switching to their conventions). However, the general asymptotic relation (\ref{largeNbeh})
provided here allows to derive analogous, new results
for {\em any} $\nu$ for which the smallest eigenvalue in the Dirac picture is known.

We have numerically verified the scaling behavior above for \textsf{FTWL} ensemble
for the two cases $\nu=1,3$ in fig. \ref{momentexpl}.
We plot the combination $N^3 \langle \lambda_{\mathrm{min}}\rangle_{N,\nu,1}$ as a function of $N$ for $\nu=1,3$
(left and right panel respectively), which nicely converge for $N\to\infty$ to the predicted values $(1/4)\kappa_{1,1,1}=2$ and
$(1/4)\kappa_{1,3,1}=e^2-1\approx 6.38906$, where we have used (\ref{kappadef}) along with (\ref{pb3}).

\section{Conclusions}\label{concl}

In this paper we addressed the smallest eigenvalue distribution in fixed-trace Wishart-Laguerre (\textsf{FTWL}) ensembles of random matrices, as well as
its cumulative distribution and moments, and the corresponding quantity for unconstrained Wishart-Laguerre (\textsf{WL}) ensembles.
Our motivation comes from the statistical description of entangled random pure states in bipartite
systems of size $MN$, where the distribution of the smallest Schmidt eigenvalue provides useful information about the degree of entanglement of these states, 
and as such it has been subject to intense scrutiny in recent years.

In the first part, we derived new results for the \textsf{FTWL} ensemble with orthogonal symmetry. Here the constraint plays an
important role, and we computed explicit expressions for the smallest
eigenvalue by Laplace inverting a recursion relation given by Edelman for the
unconstrained \textsf{WL} ensemble. In particular our results extend very recent
expressions to even values of $\nu=M-N$. Examples were given for several even and odd values of $\nu$ and checked via numerical simulations.

In the second part, we studied the microscopic large-$N$ limit at fixed $\nu$, both for orthogonal and unitary ensembles. Here the constraint (after a proper rescaling) is immaterial due to universality. We proved the equivalence of two different sets of results in the literature, given in terms of either
hypergeometric functions of matrix argument (\textsf{HFMA}), or in terms of
Pfaffians and determinants of Bessel functions. Our technical tool was to translate
\textsf{HFMA} into their multiple integral representations and to use the classical Andr\'eief-de Brujin identities.
Apart from this equivalence proof, our new results include the cumulative distributions, the universality of the smallest eigenvalue for $\nu=0$ and an explicit expression for $\nu=2$, all in the orthogonal symmetry class. In addition we have determined the precise asymptotic behavior of all moments for large $N$, in both symmetry classes.

Several open problems persist, in particular how to find a closed universal formula for all even values of $\nu$ in the orthogonal symmetry class in the microscopic limit. Furthermore, it would be interesting to extend these results to the symplectic symmetry class. Here a compact expression is only known for the special integer $\nu=0$. The problem is the apparent lack of a suitable integral representation for \textsf{HFMA} in order to proceed.
\vspace{10pt}

{\bf Acknowledgments.}
We would like to thank Satya Majumdar for discussions.

\appendix
\section{Hypergeometric functions of matrix argument (\textsf{HFMA})}\label{hypA}

The hypergeometric function of a matrix argument takes
a complex symmetric matrix $(m\times m)$ $\mathcal{X}$ as input and
provides a real number as output. More precisely, let $p\geq 0$ and $q\geq 0$ be integers, and $\{x_1,\ldots,x_m\}$ be
the real eigenvalues of $\mathcal{X}$. We have:

\begin{equation}\label{hypergeometric}
\fl  {_pF_q^{(\beta)}}(a_1,\ldots,a_p;b_1,\ldots,b_q;\mathcal{X}):=\sum_{k=0}^\infty
  \sum_{\kappa\vdash k}\frac{(a_1)_\kappa^{(\beta)}\ldots (a_p)_\kappa^{(\beta)}}{k!(b_1)_\kappa^{(\beta)}\ldots (b_q)_\kappa^{(\beta)}}
  C_\kappa^{(\beta)}(\mathcal{X}),
\end{equation}
where $\beta>0$ is a parameter, the symbol $\kappa\vdash k$ means that $\kappa=(\kappa_1,\kappa_2,\ldots)$ is a
partition of $k$ (i.e. $\kappa_1\geq\kappa_2\geq\ldots\geq 0$ are integers such that $|\kappa|=\kappa_1+\kappa_2+\ldots=k$), and the following symbol
$(a)_\kappa^{(\beta)}=\prod_{(i,j)\in\kappa}\left(a-\frac{i-1}{\beta}+j-1\right)$
is a generalized Pochhammer symbol. In (\ref{hypergeometric}), $C_\kappa^{(\beta)}(\mathcal{X})$ is a Jack polynomial, i.e.
a symmetric, homogeneous polynomial of degree $|\kappa|$ in the eigenvalues $\{x_i\}$ of $\mathcal{X}$ \cite{muir,stanley}.
Jack polynomials generalize the Schur
function, the zonal polynomial and the quaternion zonal polynomial
to which they reduce for $\beta=1,2,4$ respectively. The \textsf{HFMA} (\ref{hypergeometric}) generalizes
the ordinary hypergeometric function to which it reduces when $m=1$. 

The series (\ref{hypergeometric}) converges for any $\mathcal{X}$
if $p\leq q$; it converges if $\max_i |x_i|<1$ and $p=q+1$; and
diverges if $p>q+1$, unless it terminates. An efficient evaluation of \textsf{HFMA}
is now made possible by an algorithm devised by Koev and Edelman \cite{koev}, which we used extensively
to check numerically all the results given in terms of \textsf{HFMA} in this paper.

\section{Microscopic limit of $p^{(\mathrm{FT})}_{N,\nu=0}(x)$ for $\beta=1$}\label{nu0lim}

In this appendix we take the microscopic limit eq. (\ref{aspp}) of the smallest eigenvalue for $\beta=1$ and $\nu=0$ in \textsf{FTWL} eq. (\ref{nu0p}):
\begin{eqnarray}
\fl&&\lim_{N\to\infty}\frac{1}{4N^3}
p_{N,\nu=0}^{(\mathrm{FT})} \left(\frac{y}{4N^3}\right) =
\lim_{N\to\infty}
\frac{N\Gamma(N) \Gamma(N^2/2)}{N^32^{N+1}\Gamma(N/2)\Gamma((N^2+N-2)/2)}
{\left(\frac{y}{4N^3}\right)^{-N/2}}
\non\\
\fl&&\times{\left(1-\frac{Ny}{4N^3}\right)^{(N^2+N-4)/2}}
{_2F_1}\left(\frac{N+2}{2},\frac{N-1}{2};\frac{N^2+N-2}{2};N-\frac{4N^3}{y}\right).
\label{limnu0p}
\end{eqnarray}
While the first factor in the second line yields an exponential prefactor,
\be
\lim_{N\to\infty}{\left(1-\frac{y}{4N^2}\right)^{(N^2+N-4)/2}}\ =\ e^{-y/8}\ ,
\label{expfac}
\ee
the asymptotic limit of the hypergeometric function requires more care.
Using eq. 9.132.1 in \cite{Grad} we have that
\begin{eqnarray}
\fl&&{_2F_1}\left(\frac{N+2}{2},\frac{N-1}{2};\frac{N^2+N-2}{2};N-\frac{4N^3}{y}\right)
\non\\
\fl&=&X^{\frac{N+2}{2}}
\frac{\Gamma\left(\frac{N^2}{2}+\frac{N}{2}-1\right)\Gamma\left(-\frac32\right)}
{\Gamma\left(\frac{N}{2}-\frac12\right)\Gamma\left(\frac{N^2}{2}-2\right)}
{_2F_1}\left(\frac{N}{2}+1,\frac{N^2-1}{2};\frac52;X\right)
\non\\
\fl&&+X^{\frac{N+1}{2}}
\frac{\Gamma\left(\frac{N^2}{2}+\frac{N}{2}-1\right)\Gamma\left(\frac32\right)}
{\Gamma\left(\frac{N}{2}+1\right)\Gamma\left(\frac{N^2}{2}-\frac12\right)}
{_2F_1}\left(\frac{N-1}{2},\frac{N^2}{2}-2;-\frac12;X\right)
\label{Fid0}
\end{eqnarray}
with
\be
X=\left(\frac{4N^3}{y}-N+1\right)^{-1}.
\ee
In order to apply the standard asymptotic limits, eqs. 9.121.9-10 in \cite{Grad},
\begin{eqnarray}
\lim_{j,k\to\infty} {_2F_1}\left(j,k;\frac12;\frac{z^2}{4jk}\right)&=& \cosh(z)\\
\lim_{j,k\to\infty} {_2F_1}\left(j,k;\frac32;\frac{z^2}{4jk}\right)&=& \frac{\sinh(z)}{z}\ ,
\label{Fasymp}
\end{eqnarray}
where we identify $j=N/2$ and $k=N^2/2$, 
we still have to shift the third index of the two hypergeometric functions in eq. (\ref{Fid0}). Using eq. 9.137.1 in \cite{Grad} we obtain for the first
\begin{eqnarray}
\fl &&{_2F_1}\left(\frac{N}{2}+1,\frac{N^2-1}{2};\frac52;X\right)
\non\\
\fl&=&\frac{-1}{\left(\frac{N}{2}-\frac12\right)\left(\frac{N^2}{2}-2\right)X}
\left[\frac34(X-1)\ {_2F_1}\left(\frac{N}{2}+1,\frac{N^2-1}{2};\frac12;X\right)
\right.\non\\
\fl&&+\left. \frac32\left(\frac12+\left(\frac{N^2}{2}+\frac{N}{2}-\frac32\right)X\right)
{_2F_1}\left(\frac{N}{2}+1,\frac{N^2-1}{2};\frac32;X\right)
\right],
\label{Fid1}
\end{eqnarray}
and for the second
\begin{eqnarray}
\fl &&{_2F_1}\left(\frac{N-1}{2},\frac{N^2}{2}-2;-\frac12;X\right)
\non\\
\fl&=&\frac{-1}{\frac12\left(-\frac12\right)(X-1)}
\left[\left(\frac{N}{2}-1\right)\left(\frac{N^2}{2}-\frac52\right)X\ {_2F_1}\left(\frac{N-1}{2},\frac{N^2}{2}-2;\frac32;X\right)
\right.\non\\
\fl&&+\left. \frac12\left(-\frac12+\left(\frac{N^2}{2}+\frac{N}{2}-\frac52\right)X\right)
{_2F_1}\left(\frac{N-1}{2},\frac{N^2}{2}-2;\frac12;X\right)
\right].
\label{Fid2}
\end{eqnarray}
Collecting all the results from above and all factors we obtain as a final result
\begin{eqnarray}
\fl\lim_{N\to\infty}\frac{1}{4N^3}
p_{N,\nu=0}^{(\mathrm{FT})} \left(\frac{y}{4N^3}\right) &=&
\frac14 e^{-y/8}
\left[
\frac12\left(\cosh\left(\frac{\sqrt{y}}{2}\right)-\frac{2}{\sqrt{y}}\sinh\left(\frac{\sqrt{y}}{2}\right)\right)
\right.\non\\
\fl&&\left.+\frac{1}{\sqrt{y}}\left(\cosh\left(\frac{\sqrt{y}}{2}\right)-\frac{\sqrt{y}}{2}\sinh\left(\frac{\sqrt{y}}{2}\right)\right)
\right]
\non\\
\fl&=&\frac18 \left(1+\frac{2}{\sqrt{y}}\right)
e^{-y/8-\sqrt{y}/2}
\end{eqnarray}
which agrees with eq. (\ref{WLb1nu0lim}) as we have claimed.

\vspace{30pt}


\end{document}